\documentclass[aps,prb,twocolumn,superscriptaddress,floatfix,letterpaper]{revtex4-1}

\renewcommand\textcite[1]{\Ref{#1}}

\newcommand\Rep{\mathrm{Rep}}
\newcommand\sRep{\mathrm{sRep}}

\begin{document}

\begin{titlepage}

\title{Hierarchy construction and non-Abelian families of generic topological orders}

\author{Tian Lan} 
\affiliation{Perimeter Institute for Theoretical Physics, Waterloo, Ontario N2L 2Y5, Canada} 
\affiliation{Department of Physics and Astronomy,
  University of Waterloo, Waterloo, Ontario N2L 3G1, Canada}

\author{Xiao-Gang Wen}
\affiliation{Department of Physics, Massachusetts Institute of
Technology, Cambridge, Massachusetts 02139, USA}

\begin{abstract} 
We generalize the hierarchy construction to generic 2+1D topological orders
(which can be non-Abelian) by condensing Abelian anyons in one topological
order to construct a new one.  We show that such construction is reversible and
leads to a new equivalence relation between
topological orders. We refer to the corresponding equivalent class (the orbit of
the hierarchy construction) as ``the non-Abelian family''.  Each non-Abelian
family has one or a few root topological orders with the smallest number of
anyon types.  All the Abelian topological orders belong to the trivial
non-Abelian family whose root is the trivial topological order.  We show that
Abelian anyons in root topological orders must be bosons or
fermions with trivial mutual statistics between them.  The classification of
topological orders is then greatly simplified, by focusing on the roots of each
family: those roots are given by non-Abelian modular extensions of
representation categories of Abelian groups.

\end{abstract}


\maketitle

\end{titlepage}


\noindent
\textbf{Introduction}:
The ultimate dream of classifying objects in nature may be creating a ``table''
for them. A classic example of such classification result is the ``Periodic
Table'' for chemical elements. As for the topological ordered\cite{Wrig,KW9327}
phases of matter, which draws more and more research interests recently, we are
already able to create some ``tables'' for
them\cite{RSW0777,BBC1440,W150605768,LW150704673,LW160205936,LW160205946}, via
the theory of \mbox{(pre-)modular}
categories.  However, efforts are needed to further understand the tables, for
example, to reveal some ``periodic'' structures in the table.

In the Periodic Table, elements are divided into several ``families'' (the
columns of the table), and those in the same family have similar chemical
properties. The underlying reason for this is that elements in the same family
have similar outer electron structures, and only differ by ``noble gas cores''.
The last family consists of noble gas elements, which are chemically ``inert'',
as they have no outer electrons besides the noble gas cores. Thus the
``family'' can be considered as the equivalent class up to the ``inert'' noble
gas elements.

When it comes to topological orders, we also have ``inert'' ones: the Abelian
topological orders are ``inert'', for example, in the application of
topological quantum computation\cite{Kit03,FKLW03}. Abelian anyons can not
support non-local topological degeneracy, which is an essential difference from
non-Abelian anyons. Is it possible to define equivalent classes for topological
orders, which are up to Abelian topological orders? In this letter, we use the
hierarchy construction to establish such equivalent classes, which we will call
the ``non-Abelian families''.  The hierarchy construction is well known in the
study of Abelian fractional quantum Hall (FQH)
states\cite{TSG82,Lau83,H8305,H8483}.  In this letter we generalize it to
arbitrary (potentially non-Abelian) topological orders. 

We show that the generalized hierarchy construction is reversible. Thus, we
can say that two topological orders belong to the same ``non-Abelian family''
if they are related by the hierarchy construction. 
Each non-Abelian family has special ``root'' topological orders (see Table
\ref{roots}), with the following properties:
\begin{enumerate}
  \item Root states have the smallest rank (number of anyon types) among
    the non-Abelian family.
  \item Abelian anyons in a root state are all bosons or fermions, and have
    trivial mutual statistics with each other.
\end{enumerate}
Since any topological order in the same non-Abelian family can be reconstructed
from a root state, our work simplifies the classification of generic
topological orders to the classification of root states.

Our calculation is based on quantitative characterizations of topological
orders. One way to do so is to use the $S,T$ \emph{modular matrices} obtained
from the non-Abelian geometric phases of degenerate ground states on torus
\cite{Wrig,KW9327}. We will show,  starting from a topological order described
by $S,T$,  how to obtain another topological order described by new $S',T'$ via
a condensation of Abelian anyons. (For a less general approach based on wave
functions, see \Ref{BS07113204}.) The calculation uses the theory of fusion and
braiding of quasiparticles (which will be called \emph{anyons}) in topological
order. Such a theory is the so called ``unitary modular tensor category (UMTC) theory''
(for a review and much more details on UMTC, see \Ref{W150605768}).

A UMTC $\cC$ is simply a set of anyons (two anyons connected by a local operator
are regarded as the same type), plus data to describe their fusion and
braiding.  Like the fusion of two spin-1 particles give rise to a ``direct
sum'' of spin-0,1,2 particles: $ 1 \otimes 1 =0\oplus 1 \oplus 2$, the fusion
of two  anyons $i$ and $j$ in general gives rise to a ``direct sum'' of several
other anyons: $ i\otimes j = \bigoplus_k N^{ij}_k k$.  So the fusion of anyons
is quantitatively described by a rank-3 integer tensor $N^{ij}_k$.  From
$N^{ij}_k$, we can determine the internal degrees of freedom of anyons, which
is the so called \emph{quantum dimension}. For example, the quantum dimension
of a spin-$S$ particle is $d=2S+1$.  For an anyon $i$, its quantum dimension
$d_i$, which can be non-integer, is the largest eigenvalue of matrix $N_i$ with
$(N_i)_{kj}=N^{ij}_k$.

After knowing the fusion, the braiding of anyons can be fully determined by the
fractional part of their angular momentum $L^z$: $ s_i =\text{mod}(L^z_i,1)$.
$s_i$ is called the \emph{topological spin} (or simply spin) of the anyon $i$.
The last piece of data to characterize topological orders is the \emph{chiral
central charge} $c$, which is the number of right-moving edge modes minus the
number of left-moving edge modes.  

It turns out that two sets of data $(S,T)$ and $(N^{ij}_k,s_i)$ can fully
determine each other:
\begin{align}
T_{ij} &=\ee^{ 2\pi\ii s_i}  \del_{ij},
&
 S_{ij} &=\sum_k  \ee^{ 2\pi\ii (s_i+s_j-s_k)} \frac{N^{ij}_k d_k}{D},
\nonumber\\
 \ee^{ 2\pi\ii s_i} &=T_{ii}, 
&
N^{ij}_k &=\sum_{l} \frac{ S_{li} S_{lj} \overline{S_{lk}}}{ S_{l1} } .
\end{align}
where $D=\sqrt{\sum d_i^2}$ is the total quantum dimension.

\def\arraystretch{1.25} \setlength\tabcolsep{3pt}
\begin{table}[t] 
\caption{The low-rank root topological orders for bosonic systems. We pick only
one root for each non-Abelian family.  The rank $N$ is the number of anyon
types and $c$ the chiral central charge of the edge states. $s_i$ and $d_i$ are
the topological spin and quantum dimension of type-$i$ anyon.  The anyons
in red have trivial mutual statistics with all Abelian anyons.
Here $\zeta_n^m=\frac{\sin[\pi(m+1)/(n+2)]}{\sin[\pi/(n+2)]}$.
Some roots are the stacking of simpler ones, such as
$4_0=2_{\frac{14}{5}}\boxtimes 2_{\frac{26}{5}}$.
} 
\label{roots} 
\centering
\begin{tabular}{ |c|c|l|l| }
\hline
$N_{c}$ & $D^2$ & $s_1,s_2,\cdots$ & $d_1,d_2,\cdots$ \\ 
\hline
$1_{0}$ & $1$ & ${\red {0}}$  & ${\red {1}}$  \\
$2_{\frac{14}{5}}$ & $3.618$ & ${\red {0}},{\red {\frac{2}{5}}}$  & ${\red {1}},{\red {\zeta_{3}^{1}}}$  \\
$2_{\frac{26}{5}}$ & $3.618$ & ${\red {0}},{\red {\frac{3}{5}}}$  & ${\red {1}},{\red {\zeta_{3}^{1}}}$  \\
$3_{\frac{5}{2}}$ & $4$ & ${\red {0}},{\red {\frac{1}{2}}},\frac{5}{16}$  & ${\red {1}},{\red {1}},\zeta_{2}^{1}$  \\
$3_{\frac{8}{7}}$ & $9.295$ & ${\red {0}},{\red {\frac{6}{7}}},{\red {\frac{2}{7}}}$  & ${\red {1}},{\red {\zeta_{5}^{1}}},{\red {\zeta_{5}^{2}}}$  \\
$3_{\frac{48}{7}}$ & $9.295$ & ${\red {0}},{\red {\frac{1}{7}}},{\red {\frac{5}{7}}}$  & ${\red {1}},{\red {\zeta_{5}^{1}}},{\red {\zeta_{5}^{2}}}$  \\
$4_{0}$ & $13.09$ & ${\red {0}},{\red {\frac{2}{5}}},{\red {\frac{3}{5}}},{\red {0}}$  & ${\red {1}},{\red {\zeta_{3}^{1}}},{\red {\zeta_{3}^{1}}},{\red {\zeta_{8}^{2}}}$  \\
$4_{\frac{12}{5}}$ & $13.09$ & ${\red {0}},{\red {\frac{3}{5}}},{\red {\frac{3}{5}}},{\red {\frac{1}{5}}}$  & ${\red {1}},{\red {\zeta_{3}^{1}}},{\red {\zeta_{3}^{1}}},{\red {\zeta_{8}^{2}}}$  \\
$4_{\frac{28}{5}}$ & $13.09$ & ${\red {0}},{\red {\frac{2}{5}}},{\red {\frac{2}{5}}},{\red {\frac{4}{5}}}$  & ${\red {1}},{\red {\zeta_{3}^{1}}},{\red {\zeta_{3}^{1}}},{\red {\zeta_{8}^{2}}}$  \\
$4_{\frac{10}{3}}$ & $19.23$ & ${\red {0}},{\red {\frac{1}{3}}},{\red {\frac{2}{9}}},{\red {\frac{2}{3}}}$  & ${\red {1}},{\red {\zeta_{7}^{1}}},{\red {\zeta_{7}^{2}}},{\red {\zeta_{7}^{3}}}$  \\
$4_{\frac{14}{3}}$ & $19.23$ & ${\red {0}},{\red {\frac{2}{3}}},{\red {\frac{7}{9}}},{\red {\frac{1}{3}}}$  & ${\red {1}},{\red {\zeta_{7}^{1}}},{\red {\zeta_{7}^{2}}},{\red {\zeta_{7}^{3}}}$  \\
$5_{2}$ & $12$ & ${\red {0}},{\red {0}},\frac{1}{8},\frac{5}{8},{\red {\frac{1}{3}}}$  & ${\red {1}},{\red {1}},\zeta_{4}^{1},\zeta_{4}^{1},{\red {2}}$  \\
$5_{6}$ & $12$ & ${\red {0}},{\red {0}},\frac{7}{8},\frac{3}{8},{\red {\frac{2}{3}}}$  & ${\red {1}},{\red {1}},\zeta_{4}^{1},\zeta_{4}^{1},{\red {2}}$  \\
$5_{\frac{16}{11}}$ & $34.64$ & ${\red {0}},{\red {\frac{9}{11}}},{\red {\frac{2}{11}}},{\red {\frac{1}{11}}},{\red {\frac{6}{11}}}$  & ${\red {1}},{\red {\zeta_{9}^{1}}},{\red {\zeta_{9}^{2}}},{\red {\zeta_{9}^{3}}},{\red {\zeta_{9}^{4}}}$  \\
$5_{\frac{72}{11}}$ & $34.64$ & ${\red {0}},{\red {\frac{2}{11}}},{\red {\frac{9}{11}}},{\red {\frac{10}{11}}},{\red {\frac{5}{11}}}$  & ${\red {1}},{\red {\zeta_{9}^{1}}},{\red {\zeta_{9}^{2}}},{\red {\zeta_{9}^{3}}},{\red {\zeta_{9}^{4}}}$  \\
$5_{\frac{18}{7}}$ & $35.34$ & ${\red {0}},{\red {\frac{6}{7}}},{\red {\frac{6}{7}}},{\red {\frac{1}{7}}},{\red {\frac{3}{7}}}$  & ${\red {1}},{\red {\zeta_{5}^{2}}},{\red {\zeta_{5}^{2}}},{\red {\zeta_{12}^{2}}},{\red {\zeta_{12}^{4}}}$  \\
$5_{\frac{38}{7}}$ & $35.34$ & ${\red {0}},{\red {\frac{1}{7}}},{\red {\frac{1}{7}}},{\red {\frac{6}{7}}},{\red {\frac{4}{7}}}$  & ${\red {1}},{\red {\zeta_{5}^{2}}},{\red {\zeta_{5}^{2}}},{\red {\zeta_{12}^{2}}},{\red {\zeta_{12}^{4}}}$  \\
$6_{\frac{3}{10}}$ & $14.47$ & ${\red {0}},{\red {\frac{1}{2}}},\frac{11}{16},{\red {\frac{9}{10}}},{\red {\frac{2}{5}}},\frac{7}{80}$  & ${\red {1}},{\red {1}},\zeta_{2}^{1},{\red {\zeta_{3}^{1}}},{\red {\zeta_{3}^{1}}},\zeta_{2}^{1}\zeta_{3}^{1}$  \\
$6_{\frac{77}{10}}$ & $14.47$ & ${\red {0}},{\red {\frac{1}{2}}},\frac{5}{16},{\red {\frac{1}{10}}},{\red {\frac{3}{5}}},\frac{73}{80}$  & ${\red {1}},{\red {1}},\zeta_{2}^{1},{\red {\zeta_{3}^{1}}},{\red {\zeta_{3}^{1}}},\zeta_{2}^{1}\zeta_{3}^{1}$  \\
$6_{0}$ & $20$ & ${\red {0}},{\red {0}},{\red {\frac{1}{5}}},{\red {\frac{4}{5}}},0,\frac{1}{2}$  & ${\red {1}},{\red {1}},{\red {2}},{\red {2}},\sqrt{5},\sqrt{5}$  \\
$6_{4}$ & $20$ & ${\red {0}},{\red {0}},{\red {\frac{2}{5}}},{\red {\frac{3}{5}}},\frac{1}{4},\frac{3}{4}$  & ${\red {1}},{\red {1}},{\red {2}},{\red {2}},\sqrt{5},\sqrt{5}$  \\
$6_{\frac{58}{35}}$ & $33.63$ & ${\red {0}},{\red {\frac{2}{5}}},{\red {\frac{1}{7}}},{\red {\frac{5}{7}}},{\red {\frac{19}{35}}},{\red {\frac{4}{35}}}$  & ${\red {1}},{\red {\zeta_{3}^{1}}},{\red {\zeta_{5}^{1}}},{\red {\zeta_{5}^{2}}},{\red {\zeta_{3}^{1}\zeta_{5}^{1}}},{\red {\zeta_{5}^{2}\zeta_{3}^{1}}}$  \\
$6_{\frac{138}{35}}$ & $33.63$ & ${\red {0}},{\red {\frac{2}{5}}},{\red {\frac{6}{7}}},{\red {\frac{2}{7}}},{\red {\frac{9}{35}}},{\red {\frac{24}{35}}}$  & ${\red {1}},{\red {\zeta_{3}^{1}}},{\red {\zeta_{5}^{1}}},{\red {\zeta_{5}^{2}}},{\red {\zeta_{3}^{1}\zeta_{5}^{1}}},{\red {\zeta_{5}^{2}\zeta_{3}^{1}}}$  \\
$6_{\frac{142}{35}}$ & $33.63$ & ${\red {0}},{\red {\frac{3}{5}}},{\red {\frac{1}{7}}},{\red {\frac{5}{7}}},{\red {\frac{26}{35}}},{\red {\frac{11}{35}}}$  & ${\red {1}},{\red {\zeta_{3}^{1}}},{\red {\zeta_{5}^{1}}},{\red {\zeta_{5}^{2}}},{\red {\zeta_{3}^{1}\zeta_{5}^{1}}},{\red {\zeta_{5}^{2}\zeta_{3}^{1}}}$  \\
$6_{\frac{222}{35}}$ & $33.63$ & ${\red {0}},{\red {\frac{3}{5}}},{\red {\frac{6}{7}}},{\red {\frac{2}{7}}},{\red {\frac{16}{35}}},{\red {\frac{31}{35}}}$  & ${\red {1}},{\red {\zeta_{3}^{1}}},{\red {\zeta_{5}^{1}}},{\red {\zeta_{5}^{2}}},{\red {\zeta_{3}^{1}\zeta_{5}^{1}}},{\red {\zeta_{5}^{2}\zeta_{3}^{1}}}$  \\
$6_{\frac{46}{13}}$ & $56.74$ & ${\red {0}},{\red {\frac{4}{13}}},{\red {\frac{2}{13}}},{\red {\frac{7}{13}}},{\red {\frac{6}{13}}},{\red {\frac{12}{13}}}$  & ${\red {1}},{\red {\zeta_{11}^{1}}},{\red {\zeta_{11}^{2}}},{\red {\zeta_{11}^{3}}},{\red {\zeta_{11}^{4}}},{\red {\zeta_{11}^{5}}}$  \\
$6_{\frac{58}{13}}$ & $56.74$ & ${\red {0}},{\red {\frac{9}{13}}},{\red {\frac{11}{13}}},{\red {\frac{6}{13}}},{\red {\frac{7}{13}}},{\red {\frac{1}{13}}}$  & ${\red {1}},{\red {\zeta_{11}^{1}}},{\red {\zeta_{11}^{2}}},{\red {\zeta_{11}^{3}}},{\red {\zeta_{11}^{4}}},{\red {\zeta_{11}^{5}}}$  \\
$6_{\frac{8}{3}}$ & $74.61$ & ${\red {0}},{\red {\frac{1}{9}}},{\red {\frac{1}{9}}},{\red {\frac{1}{9}}},{\red {\frac{1}{3}}},{\red {\frac{2}{3}}}$  & ${\red {1}},{\red {\zeta_{7}^{3}}},{\red {\zeta_{7}^{3}}},{\red {\zeta_{7}^{3}}},{\red {\zeta_{16}^{4}}},{\red {\zeta_{16}^{6}}}$  \\
$6_{\frac{16}{3}}$ & $74.61$ & ${\red {0}},{\red {\frac{8}{9}}},{\red {\frac{8}{9}}},{\red {\frac{8}{9}}},{\red {\frac{2}{3}}},{\red {\frac{1}{3}}}$  & ${\red {1}},{\red {\zeta_{7}^{3}}},{\red {\zeta_{7}^{3}}},{\red {\zeta_{7}^{3}}},{\red {\zeta_{16}^{4}}},{\red {\zeta_{16}^{6}}}$  \\
$6_{2}$ & $100.6$ & ${\red {0}},{\red {\frac{6}{7}}},{\red {\frac{5}{7}}},{\red {\frac{3}{7}}},{\red {0}},{\red {\frac{1}{3}}}$  &  ${\red {1}},{\red {\frac{3+\sqrt{21}}{2}}\times 3},{\red {\frac{5+\sqrt{21}}{2}}},{\red {\frac{7+\sqrt{21}}{2}}}$  \\
$6_{6}$ & $100.6$ & ${\red {0}},{\red {\frac{1}{7}}},{\red {\frac{2}{7}}},{\red {\frac{4}{7}}},{\red {0}},{\red {\frac{2}{3}}}$  & ${\red {1}},{\red {\frac{3+\sqrt{21}}{2}}\times 3},{\red {\frac{5+\sqrt{21}}{2}}},{\red {\frac{7+\sqrt{21}}{2}}}$  \\
\hline
\end{tabular}
\end{table}

~

\noindent
\textbf{Hierarchy construction in generic topological orders}: Let us consider
an Abelian anyon condensation in a generic topological order, described by a
UMTC $\cC$.  (Such a condensation in an Abelian topological order is
discussed in Appendix \ref{Hab}.) The anyons in $\cC$ are labeled by
$i,j,k,\cdots$.  Let $a_c$ be an Abelian anyon in $\cC$ with spin $s_c$.  We
condense $a_c$ into the Laughlin state $ \Psi=\prod(z_i-z_j)^{m_c-2s_c}$,
where $m_c=$ even and $m_c-2s_c\neq 0$.  \footnote{This is different from the
so called ``anyon condensation'' categorical approach where only bosons
condensing into the trivial state is considered.} The resulting topological
order is described by UMTC $\cD$, determined by $\cC$, $a_c$ and $m_c$.  

To calculate $\cD$, we note that the anyons in $\cD$ are the anyons $i$ in
$\cC$ dressed with the vortices of the Laughlin state of $a_c$.  The vorticity
is given by $m-t_i$, where $m$ is an integer, and $2\pi t_i$ is the mutual
statistics angle between anyon $i$ and the condensing anyon $a_c$ in the
original topological order $\cC$, which can be extracted from the $S$ matrix
$\ee^{-2\pi\ii t_i}=S^\cC_{i a_c}/S^\cC_{i\one}$, or $-t_i = s_i + s_{a_c} -
s_{i\otimes a_c}$.  Thus anyons in $\cD$ are labeled by pairs $I=(i,m)$.  We
like to ask what is the spin and fusion rules of $I=(i,m)$?  

The spin of $(i,m)$ is given by the spin of $i$ plus the spin of the $m-t_i$
flux in the Laughlin state:
\begin{align}
 s_{(i,m)}  & = s_i +\frac12 \frac{(m-t_i)^2 }{m_c-2s_c}
\label{spinim}
\end{align}
Fusing $i$ with $m-t_i$ flux and $j$ with $n-t_j$ flux gives us
$i\otimes j$ with $m-t_i + n-t_j$ flux:
\begin{align}
 (i,m)&\otimes(j,n) \sim \oplus_{k}N^{ij}_k(k,m-t_i+n-t_j+t_k),
\label{fusionim}
\end{align}
where $N^{ij}_k$ is the fusion coefficient in $\cC$.
Since $a_c$ with $m_c-t_{a_c}=m_c-2s_c$ flux is condensed, fusing with
$(a_c,m_c)$ anyon does not change the anyon type in $\cD$.  So, we have
an equivalence relation:
\begin{align}
 (i,m) & \sim (i\otimes a_c, m-t_i+m_c-2s_c+t_{i\otimes a_c}),
\label{equivim}
\end{align}
The above three relations fully determine the topological order $\cD$.\cite{W150605768,LW150704673}

It is important to fix a ``gauge'' for $t_i$, say by choosing $t_i
\in [0,1)$.  The same label $(i,m)$ may label different anyons under different
``gauge'' choices of $t_i$. Similarly, we have fixed a ``gauge'' for $s_c$ that
fixed the meaning of $m_c$.  Note that $t_{a_c}$ is automatically fixed when
$s_c$ is fixed: $  t_{a_c}=2 s_c$, while other $t_i$ can be freely chosen. This
ensures that the equivalence relation \eqref{equivim} is compatible with fusion
\eqref{fusionim}, where \eqref{equivim} is generated by fusing with the
trivial anyon $(a_c,m_c)$.  The combinations $m-t_i$,
$m_c-2s_c$ determine the final spins and fusion rules; they are
gauge-invariant quantities. Thus, if we change the gauge of $t_i,s_c$, i.e.,
modify them by some integers, $m,m_c$ should be modified by the same integers to
ensure that the construction remains the same.

Below we will study the properties of $\cD$ in detail. Let $M_c=m_c-2s_c$.
Applying the equivalence relation \eqref{equivim} $q$ times, we obtain
\begin{align}\label{equivq}
  (i,m)\sim (i\otimes a_c^{\otimes q},m-t_i+qM_c+t_{i\otimes a_c^{\otimes
  q}}).
\end{align}
Let $q_c$ be the ``period'' of $a_c$, i.e., the smallest positive integer such
that $a_c^{\otimes q_c}=\one$. We see that
\begin{align}
  (i,m)\sim (i,m+q_cM_c).
\end{align}
Thus, we can focus on the reduced range of $m\in\{0,1,2,\cdots,q_c|M_c|-1\}$.
Let $|\cC|, |\cD|$ denote the rank of $\cC, \cD$ respectively. Now within the
reduce range of $m$, we have $q_c|M_c||\cC|$ different labels, and we want to
show that the orbit generated by the equivalence relation \eqref{equivq} all
have the same length, which is $q_c$. To see this, just note that for
$0<q<q_c$, either $i\neq i\otimes a_c^{\otimes q}$, or if $i=i\otimes
a_c^{\otimes q}$, $m\neq m-t_i+qM_c+t_{i\otimes a_c^{\otimes q}}=m+qM_c$; in
other words, the labels $(i,m)$ are all different within $q_c$ steps. It
follows that the rank of $\cD$ is $|\cD|=|M_c||\cC|$.  

Strictly speaking, anyons in $\cD$ should one-to-one correspond to the
equivalent classes of $(i,m)$. However, as the orbits have the same length, it
would be more convenient to use $(i,m)$ directly (as we will see later, this
is the same as working in a pre-modular category $\tilde\cD$). For example, when we need to
sum over anyons in $\cD$, we can instead do
$
  \sum_{I\in\cD}\to\frac{1}{q_c}\sum_{i\in\cC}\sum_{m=0}^{q_c|M_c|-1}
$.
Now we are ready to calculate other quantities of the new topological order
$\cD$. First, it is easy to see that the quantum dimensions remain the same
$  d_{(i,m)}=d_i$.
The total quantum dimension is then
\begin{align}
  D_\cD^2=\frac{1}{q_c}\sum_{i\in\cC}\sum_{m=0}^{q_c|M_c|-1}
  d_{(i,m)}^2=|M_c|D_\cC^2.
\end{align}
The $S$ matrix is
\begin{align}
\label{Sd}
  S^{\cD}_{(i,m),(j,n)}&=\sum_k \frac{N^{ij}_k}{D_\cD} d_k \ee^{2\pi\ii
    [s_{(i,m)}+s_{(j,n)}-s_{(k,m+n+t_k-t_i-t_j)}]}\nonumber\\
    &=\frac{1}{\sqrt{|M_c|}}S^{\cC}_{ij}\ee^{-2\pi\ii\frac{(m-t_i)(n-t_j)}{M_c}}.
\end{align}
It is straightforward to check that $S^{\cD}_{(i,m),(j,n)}$ is unitary (with
respect to equivalent classes of $(i,m)$). Moreover, this formula for $S$ can
recover the equivalence relation \eqref{equivq} and fusion rules
\eqref{fusionim} via unitarity and Verlinde formula.

The new $S^\cD,T^\cD$ matrices ($T^\cD$-matrix is determined by the spin of
anyons $s^\cD_{(i,m)}$ in \eqref{spinim}), as well as $S^\cC,T^\cC$, should
both obey the modular relations $STS=\ee^{2\pi\ii \frac{c}{8}}T^\dag S T^\dag$,
from which we can extract the central charge of $\cD$. The new
central charge is found to be (see Appendix \ref{appdiffc})
\begin{align}
  c^\cD=c^\cC+\sgn(M_c) .\label{diffc}
\end{align}
Clearly, the one-step hierarchy construction described by \eqref{spinim},
\eqref{Sd}, and \eqref{diffc} is fully determined by an Abelian anyon $a_c$ and
$M_c$, where $M_c+2s_c$ is an even integer.  In Appendix \ref{calFR},  we
discuss the above hierarchy construction more rigorously at the full
categorical level.

As an application, let us explain  the ``eight-fold way'' observed in the table
of topological orders\cite{W150605768,LW150704673}: whenever there is a
fermionic quasiparticle, the topological order has eight companions with the
same rank and quantum dimensions but different spins and central charges. If we
apply the one-step condensation with $a_c$ being a fermion, and $M_c=\pm 1$, a
new topological order of the same rank is obtained.  \footnote{Physically this
amounts to condensing the fermionic quasiparticle into an integer quantum Hall
state. If we instead condense the fermionic quasiparticle into $p\pm\ii p$
states we are able to obtain the ``sixteen fold way''.  However, such
condensation is beyond the construction of this work.} The spins of the anyons
carrying fermion parity flux (having non-trivial mutual statistics with the
fermion $a_c$) are shifted by $\pm1/8$, and the central charge is shifted by
$\pm 1$, while all the quantum dimensions remain the same.  If we repeat it
eight times, we will go back to the original state (up to an $E_8$ state), generating the
``eight fold way''.

~

\noindent
\textbf{Reverse construction and non-Abelian families}:
The one-step condensation from $\cC$ to $\cD$ is always reversible. In $\cD$,
choosing $a_c'=(\one,1),\ s_c'=\frac{1}{2M_c},\ m_c'=0,\ M_c'=-1/M_c$, and repeating
the construction, we will go back to $\cC$.  
One may first perform the construction for a pre-modular $\tilde\cD$ and then
reduce the resulting category to a modular category. Taking
$(j,n)=\overline{a_c'}=(\one,-1)$ in \eqref{Sd} we find that the mutual
statistics between $(i,m)$ and $a_c'=(\one,1)$ is
$t'_{(i,m)}=\frac{m-t_i}{M_c}$.
Let $(i,m,p),(j,n,q)$ label the anyons after the above one-step condensation;
the new $S$ matrix is
\begin{align}
  &S_{(i,m,p),(j,n,q)}=S^\cC_{ij}\ee^{-\frac{2\pi\ii(m-t_i)(n-t_j)}{M_c}}\ee^{-\frac{2\pi\ii(p-t'_{(i,m)})(q-t_{(j,n)}')}{M'_c}}
  \nonumber\\
  &\ \ =S^\cC_{ij}\ee^{2\pi\ii (t_i q+ t_j p -t_{a_c} pq)}
  =S^\cC_{i\otimes \bar a_c^{\otimes p},j\otimes \bar a_c^{\otimes q}},
\end{align}
which means that we can identify $(i,m,p)$ with $i\otimes \bar a_c^{\otimes p}$
($\bar a_c$ denotes the anti-particle of $a_c$).
It is easy to check that they have the same spin $s_{(i,m,p)}=s_{i\otimes \bar
  a_c^{\otimes p}}$. Therefore, $i\sim (i\otimes a_c^{\otimes p},m,p),\ \forall
  m,p$, we have come back to the original state $\cC$.
Therefore, generic hierarchy constructions are reversible, which defines an
equivalence relation between topological orders. We call the corresponding
equivalent classes the ``non-Abelian families''.

Now we examine the important quantity $M_c=m_c-2s_c$ which relates the ranks
before and after the one-step condensation, $|\cD|=|M_c||\cC|$. Since $m_c$ is
a freely chosen even integer, when $a_c$ is not a boson or fermion ($s_c\neq 0$
or $1/2 \mod 1$), we can always make $0<|M_c|<1$, which means that the rank is
reduced after one-step condensation. We then have the first important
conclusion: \emph{Each non-Abelian family have ``root'' topological orders with
the smallest rank; the Abelian anyons in the ``root'' states are all bosons or
fermions}.

We can further show that the Abelian bosons or
fermions in the ``root'' states have trivial mutual statistics among them.
To see this, assuming that $a,b$ are Abelian anyons in a root state. Since
the mutual statistics is given by $DS_{ab}=\exp[2\pi\ii(s_a+s_b-s_{a\otimes
b})]$, and $a,b,a\otimes b$ are all bosons or fermions, non-trivial mutual
statistics can only be $DS_{ab}=-1$. Now consider two cases: (1) one of $a,b$,
say $a$, is a fermion, then by condensing $a$ (choosing $a_c=a$,
$m_c=2$, $s_c=1/2$, $t_b=1/2$), in the new topological order, the rank remains the same but
$s_{(b,0)}=s_b+\frac{t_b^2}{2M_c}=s_b+1/8$, which means $(b,0)$ is an Abelian
anyon but neither a boson nor a fermion. By condensing $(b,0)$ again we can
reduce the rank, which conflicts with the ``root'' state assumption. (2) $a,b$
are all bosons. Still we condense $a$ with $m_c=2,s_c=0,t_b=1/2$. In the new
topological order the rank is doubled but
$s_{(b,0)}=s_b+\frac{t_b^2}{2M_c}=1/16$, which means further condensing $(b,0)$
with $m_c'=0$ the rank is reduced to $1/8$, which is again, smaller than the
rank of the beginning root state, thus contradictory.

Therefore, \emph{in the root states, Abelian anyons are bosons or fermions with
trivial mutual statistics}.
We also have a straightforward corollary: \emph{all Abelian topological orders
have the same unique root state, which is the trivial topological order}.  In
other words, all Abelian topological orders are in the same trivial non-Abelian
family, which resembles the noble gas family in the Periodic Table. Thus, the
non-Abelian families are indeed equivalent classes up to Abelian topological
orders.

To easily determine if two states belong to the same non-Abelian family, 
it is very helpful to introduce some \emph{non-Abelian invariants}. One is the fractional part of
the central charge.  Since the one-step condensation changes the central charge
by $\sgn(M_c)$ (see \eqref{diffc}), we know that central charges in the same
non-Abelian family can only differ by integers.
Another invariant is the quantum dimension. It is not hard to check that, in
the one-step condensation, the number of anyons with the same quantum dimension
is also multiplied by $|M_c|$. The third invariant is a bit involved. Note that
in the one-step condensation, if $i$ has trivial mutual statistics with $a_c$,
$t_i=0$, then $(i,0)$ in $\cD$ have the same spin as $i$ in $\cC$ and the same
mutual statistics with $(j,m),\forall m$ as that between $i$ and $j$ in $\cC$.
Therefore, the centralizer of Abelian anyons, namely, the subset of anyons that
have trivial mutual statistics with all Abelian anyons (the anyons in red in
Table \ref{roots}), is the same within one non-Abelian family. These facts
enable us to quickly tell that two states are not in the same non-Abelian
family.

~

\noindent \textbf{Examples}: Realizations of non-Abelian FQH states were first
proposed in \textcite{Wnab,MR9162}. One of them 
is \cite{Wnab,W9927}
\begin{align}
\Psi_{\nu=1}(\{z_i\})=[\chi_2(\{z_i\})]^2, 
\end{align}
where $\chi_k(\{z_i\})$ is the many-fermion wave function with $k$ filled
Landau levels.  The bulk effective theory is the
$SU(2)_{-2}^f$ Chern-Simons (CS) theory with 3 types of anyons and the edge has 
$c=5/2$ (see Appendix \ref{tables}). So the state is one of the root state $N_c=3_{\frac 52}$ in Table
\ref{roots}.  Another bosonic non-abelian FQH liquid at $\nu=1$ is
\cite{MR9162}
\begin{align} 
\Psi_{\nu=1} = \cA(\frac{1}{z_1-z_2} \frac{1}{z_2-z_3} \cdots)\prod (z_i-z_j), 
\end{align} 
whose edge has a  chiral central charge $c=3/2$.  It is the state described by
$N_c=3_{\frac 32}$ which belong to the same non-Abelian family as the $3_{\frac
52}$ state above.  The experimentally realized $\nu=5/2$ FQH state is likely to
belong to this non-Abelian family \cite{WES8776,DM08020930,RMM0899}.  

A more interesting non-Abelian state (which can perform universal topological quantum computation \cite{FLZ0205}) is
\begin{align}
\Psi_{\nu=\frac32}(\{z_i\})=[\chi_3(\{z_i\})]^2, 
\end{align}
whose edge has a  chiral central charge  $c=\frac{21}{5}$. The bulk effective
theory is the $SU(2)_{-3}^f$ CS theory with 4 types of anyons \cite{Wnab,W9927}.
So the state is $N_c=4_{\frac {21}{5}}$, which belongs to the same non-Abelian
family as the state $2_{\frac {26}{5}}$ in Table \ref{roots} (see Appendix
\ref{tables}, which contains more examples of non-Abelian states and
non-Abelian families).




We like to remark that the topological orders studied in this paper do not
require and do not have any symmetry. However, some $c=0$ topological orders
with a $Z_2$ automorphism $i\to i'$ that changes the sign of spins
$s_i =-s_{i'}$ can be realized by
time-reversal symmetric states \cite{BW161207792}.

~

\noindent
\textbf{Conclusion and Outlook}:
In this letter we introduced the hierarchy construction in generic topological
orders, which established a new equivalence relation: Two topological orders
related by the hierarchy construction belong to the same ``non-Abelian
family''. This reveals intriguing new structures in the classification of
topological orders.
Non-Abelian families are equivalent classes up to Abelian topological orders.
Topological orders in the same non-Abelian family share some properties, such
as quantum dimensions and the fractional part of central charges.

In particular we studied the ``root'' states, the states in a non-Abelian
family with the smallest rank.  Other states can be constructed from the root
states via the hierarchy construction.  Thus, classifying all topological
orders is the same as classifying all root states, namely, all states such that
their Abelian anyons have trivial mutual statistics. In other words, we can try
to generate all possible topological orders by constructing all the root
states, which can be obtained by \emph{starting with an Abelian group $G$,
extending its representation category $\Rep(G)$ or $\sRep(G^f)$ to a modular
category\cite{LW160205936,LW160205946} while requiring all the extra anyons
being non-Abelian (which is referred to as a non-Abelian modular extension).}
This is a promising future problem and may be an efficient way to produce
tables of topological orders. 

Although in this letter we focused on bosonic topological orders with no
symmetry (described by modular categories), the construction also applies to
bosonic/fermionic topological orders with any symmetry (described by certain
pre-modular categories)\cite{LW160205936,LW160205946}. The same argument goes
for non-Abelian families and root states with symmetries.  


TL thanks Zhenghan Wang for helpful discussions. This research was supported by NSF Grant No.  DMR-1506475 and NSFC 11274192.

\bibliography{local,../../bib/wencross,../../bib/all,../../bib/publst} 

\appendix

\section{Hierarchy construction in Abelian topological orders}
\label{Hab}

In this section, we will discuss hierarchy construction, \ie Abelian anyon
condensation, in Abelian topological orders in a very general setting.  This
motivates the similar construction for generic non-Abelian states discussed in
the main text.

Consider a bosonic Abelian topological order, which can always be described by
an even $K$-matrix $K_0$ of dimension $\ka$.  Anyons are labeled by
$\ka$-dimensional integer vectors $\v l_0$.  Two
integer vectors $\v l_0$ and $\v l'_0$ are equivalent (\ie describe the same
type of topological excitation) if they are related by
\begin{align}
  \v l'_0 = \v l_0 + K_0 \v k,\label{eq0}
\end{align}
where $\v k$ is an arbitrary integer vector.  The mutual statistical angle between two
anyons, $\v l_0$ and $\v k_0$, is given by
\begin{align}
 \th_{\v l_0,\v k_0} =2\pi \v k_0^T K_0^{-1} \v l_0.
\end{align}
The spin of the anyon $\v l_0$ is given by
\begin{align}
 s_{\v l_0} = \frac12 \v l_0^T K_0^{-1} \v l_0.
\end{align}

In the hierarchy construction of a new topological order from an old one, a
basic step is to condense Abelian anyons into a Laughlin-like
state. Let us construct a new topological order from the $K_0$
topological order by assuming Abelian anyons labeled by $\v l_c$ condense.
Here we treat the anyon as a bound state between a boson and flux.  We then
smear the flux such that it behaves like an additional uniform magnetic field,
and condense the boson into $\nu=1/m_c$ Laughlin state (where $m_c=$ even).
The resulting new topological order is described by the $(\ka+1)$-dimensional
$K$-matrix
\begin{align}
K_1= \bpm K_0 & \v l_c \\
     \v l_c^T &  m_c \\
\epm
\end{align}

In the following, we are going to show that, to describe the result of the $\v
l_c$ anyon condensation, we do not need to know $K_0$ directly.  We only need
to know the spin of the condensing particle $\v l_c$
\begin{align}
s_c=\frac12 \v l_c^T K_0^{-1} \v l_c,
\end{align}
and the mutual statistics 
\begin{align}
\th_{\v l_0,\v l_c}\equiv 2\pi t_{\v l_0},\ \ \  t_{\v l_0} = \v l_c^T K_0^{-1} \v l_0
\end{align}
between $\v l_0$ and  $\v l_c$.

First, we find that, as long as $m_c-2s_c\neq 0$, $K_1$ is invertible with
\begin{align}
 K_1^{-1} = 
\bpm
 K_0^{-1} +\frac{ K_0^{-1} \v l_c \v l_c^T K_0^{-1}}{ m_c - 2s_c}
& - \frac{K_0^{-1} \v l_c}{ m_c - 2s_c} \\
- \frac{\v l_c^T K_0^{-1} }{ m_c - 2s_c} &  \frac{1}{ m_c - 2s_c}
\epm
\end{align}
The anyons in the new $K_1$ topological order are labeled by
$\ka+1$-dimensional integer vector $\v l^T = (\v l_0^T, m)$.  The spin of $\v
l$ is
\begin{align}
\label{slsl0}
 s_{\v l}  &= \frac12 \v l^T K_1^{-1} \v l
 = \frac12 \Big(2s_0 + \frac{m^2+t_{\v l_0}^2 -2 m t_{\v l_0} }{m_c-2s_c}  \Big) 
\nonumber\\
&= s_{\v l_0} +\frac12 \frac{(m-t_{\v l_0})^2 }{m_c-2s_c}
\end{align}
The vectors 
$\v l^T = (\v l_0^T, m)$
and
$\v l^{\prime T} = (\v l_0^{\prime T}, m')$
are equivalent if they are related by
\begin{align}
  \label{equivk0k}
\v l_0^{\prime } - \v l_0 = K_0 \v k_0 + k \v l_c, \ \ \
m' - m = \v l_c^T \cdot \v k_0+m_c k,
\end{align}
for any $\ka$-dimensional integer vector $\v k_0$ and integer $k$. 
To avoid the gauge ambiguity, for the
integer vectors $\v l_0$, we pick a representative for each equivalent class
(by \eqref{eq0}, fixing the gauge).
Taking $k=1$ and appropriate $\v k_0$ such that $\v l_0'$ and $\v l_0$ are
the pre-fixed representatives, we see that 
\begin{align}
\label{equiv}
(\v l_0^T, m) \sim ({\v{l}'}_0^T\sim\v l_0^{T}+\v l_c^T, m+t_{\v l'_0}-t_{\v
l_0}+m_c-2s_c).
\end{align}

We also want to express the fusion in the new state in terms of the pre-fixed
representatives $\v l_1,\v l_2,\v l_3$. Assuming that $(\v l_3^T,m_3)\sim (\v
l_1^T+\v l_2^T,m_1+m_2)$, and taking $k=0$ and appropriate $\v k_0$ in
\eqref{equivk0k} (the cases of non-zero $k$ can be generated via
\eqref{equiv}), we find that
\begin{align}
  \label{fusion}
&(\v l_1^T,m_1)+(\v l_2^T,m_2)
\\
&\sim (\v l_3^T\sim \v l_1^T+\v
l_2^T,m_3=m_1+m_2+t_{\v l_3}-t_{\v l_1}-t_{\v l_2}).
\nonumber 
\end{align}

We can easily calculate the determinant of $K_1$ whose absolute value is the
rank of the new state:
\begin{align}
  \det(K_1)&=\det\bpm K_0 & \v l_c \\
     \v l_c^T &  m_c \\
     \epm = \det(K_0)(m_c-\v l_c^T K_0^{-1} \v l_c)
     \nonumber\\
&=(m_c-2s_c)\det(K_0)
\end{align}
Let $M_c=m_c-2s_c$. It is an important gauge invariant quantity relating the
ranks of the two states. If we perform the condensation with a different anyon $\v l_c'$
and a different even integer $m_c'$, but make sure that $\v l_c'\sim \v l_c$
and $M_c'=m_c'-2s_c'=m_c-2s_c=M_c$, the new topological order will be the same.

It is worth mentioning that such construction is reversible: for the $K_1$
state, take $\v l_c'^T=(\v 0^T,1),m_c'=0$, and repeat the construction:
\begin{align}
  K_2=\bpm K_0&\v l_c & 0\\
  \v l_c^T& m_c&1\\
  0&1&0\epm \sim 
  \bpm K_0& 0 & 0\\
  0& 0&1\\
  0&1&0\epm
  \sim K_0.
\end{align}
We return to the original $K_0$ state.

\section{Calculating the central charge difference of one-step condensation}
\label{appdiffc}

In the one-step condensation from $\cC$ to $\cD$, the central charge is
changed by $\sgn(M_c)$. In this section we give the detailed calculation.
Firstly, using the
modular relation for both $\cC$ and $\cD$, we find that 
\begin{align}
  \frac{1}{q_c\sqrt{|M_c|}}&\sum_{i,j,k\in\cC}\sum_{p=0}^{q_c|M_c|-1}\left\{\overline{S^\cC_{xi}}S^\cC_{ik}T^\cC_{kk}S^\cC_{kj}\overline{S^\cC_{jy}}\right.
  \nonumber\\
  &\left.\times\exp\left[\frac{2\pi\ii}{2M_c}(t_i+t_j-t_k+p)^2\right]\right\}
  \nonumber\\
  &=\exp\left( 2\pi\ii\frac{c^\cD-c^\cC}{8} \right)T^\cC_{xx}\delta_{xy}.
  \label{difc}
\end{align}
To show 
$  c^\cD-c^\cC=\sgn(M_c) \mod 8$,
we need to use the reciprocity theorem for generalized
Gauss sums\cite{BEW}:
\begin{align}
  \label{reci}
  \sum_{n=0}^{|c|-1}\ee^{\pi\ii\frac{an^2+bn}{c}}=\sqrt{|c/a|}\ee^{\frac{\pi\ii}{4}[\sgn(ac)-\frac{b^2}{ac}]}\sum_{n=0}^{|a|-1}\ee^{-\pi\ii\frac{cn^2+bn}{a}},
\end{align}
where $a,b,c$ are integers, $ac\neq 0$ and $ac+b$ even. Thus,
\begin{align}
  &\ \ \ \
\sum_{p=0}^{q_c|M_c|-1}\exp\left[\frac{2\pi\ii}{2M_c}(t_i+t_j-t_k+p)^2\right]
  \\
  &=\frac{1}{q_c}\ee^{\frac{\pi\ii(t_i+t_j-t_k)^2}{M_c}}\sum_{p=0}^{q_c^2|M_c|-1}\ee^{\frac{\pi\ii}{M_cq_c^2}\left[q_c^2p^2+2q_c^2(t_i+t_j-t_k)p\right]}
  \nonumber\\
  &=\frac{\sqrt{|M_c|}}{q_c}\ee^{\frac{\pi\ii}{4}\sgn(M_c)}\sum_{p=0}^{q_c^2-1}\ee^{-\pi\ii[M_cp^2+2(t_i+t_j-t_k)p]}
  \nonumber\\
  &=\frac{\sqrt{|M_c|}}{q_c}\ee^{\frac{\pi\ii}{4}\sgn(M_c)}\sum_{p=0}^{q_c^2-1}\ee^{-\pi\ii
  (m_c-2s_c)p^2}\frac{S^\cC_{ia_c^{\otimes p}}}{S^\cC_{i\one}}
\frac{S^\cC_{ja_c^{\otimes
p}}}{S^\cC_{j\one}}\frac{\overline{S^\cC_{ka_c^{\otimes p}}}}{S^\cC_{k\one}}
\nonumber\\
  &=\frac{\sqrt{|M_c|}}{q_c}\ee^{\frac{\pi\ii}{4}\sgn(M_c)}\sum_{p=0}^{q_c^2-1}T^\cC_{a_c^{\otimes
  p},a_c^{\otimes p}}\frac{S^\cC_{ia_c^{\otimes p}}}{S^\cC_{i\one}}
\frac{S^\cC_{ja_c^{\otimes
p}}}{S^\cC_{j\one}}\frac{\overline{S^\cC_{ka_c^{\otimes p}}}}{S^\cC_{k\one}}.
\nonumber 
\end{align}
Substituting the above result into \eqref{difc}, we have
\begin{align}
  &\frac{1}{q_c\sqrt{|M_c|}}\sum_{i,j,k\in\cC}\sum_{p=0}^{q_c|M_c|-1}\left\{\overline{S^\cC_{xi}}S^\cC_{ik}T^\cC_{kk}S^\cC_{kj}\overline{S^\cC_{jy}}\right.
  \nonumber\\
  &\ \ \ \
\left.\times\exp\left[\frac{2\pi\ii}{2M_c}(t_i+t_j-t_k+p)^2\right]\right\}
  \nonumber\\
  &=\frac{1}{q_c^2}\ee^{\frac{\pi\ii}{4}\sgn(M_c)}\sum_{p=0}^{q_c^2-1}
  \sum_{k}T_{kk}^\cC T^\cC_{a_c^{\otimes p},a_c^{\otimes p}}
  \frac{\overline{S^\cC_{ka_c^{\otimes p}}}}{S^\cC_{k\one}}
  \nonumber\\
  &\ \ \ \
\times\sum_i \frac{\overline{S^\cC_{xi}}S^\cC_{ik}S^\cC_{ia_c^{\otimes
  p}}}{S^\cC_{i\one}}\sum_j
  \frac{S^\cC_{kj}\overline{S^\cC_{jy}}S^\cC_{ja_c^{\otimes p}}}{S^\cC_{j\one}}
  \nonumber\\
  &=\frac{1}{q_c^2}\ee^{\frac{\pi\ii}{4}\sgn(M_c)}\sum_{p=0}^{q_c^2-1}
  \sum_{k}T_{k\otimes a_c^{\otimes p},k\otimes a_c^{\otimes p}}^\cC 
  N^{k,a_c^{\otimes p}}_x N^{k,a_c^{\otimes p}}_y
  \nonumber\\
  &=\frac{1}{q_c^2}\ee^{\frac{\pi\ii}{4}\sgn(M_c)}\sum_{p=0}^{q_c^2-1}
  \sum_{k}T_{k\otimes a_c^{\otimes p},k\otimes a_c^{\otimes p}}^\cC 
  \delta_{k\otimes a_c^{\otimes p},x}\delta_{xy}
  \nonumber\\
  &=\ee^{\frac{\pi\ii}{4}\sgn(M_c)}T^\cC_{xx}\delta_{xy},
\end{align}
as desired.
In fact, based on the physical picture, we have a stronger result
\begin{align}
  c^\cD=c^\cC+\sgn(M_c) .
\end{align}

So the central charge is changed by $\sgn(M_c)$ after the one-step
condensation.  A direct corollary is that the central charge of an Abelian
topological order is given by the signature of its $K$-matrix (the number of
positive eigenvalues minus the number of negative eigenvalues).

\section{The generalized hierarchy construction
at full categorical level}
\label{calFR}

Does the generalized hierarchy construction from $\cC$ to $\cD$ described by
\eqref{spinim}, \eqref{Sd}, and \eqref{diffc} always give a valid topological
order $\cD$?  To confirm this, below we will give a more rigorous construction
at full categorical level, which goes down to the level of $F,R$ matrices.

The first step is to construct a pre-modular category $\tilde\cD$, based on the
observation that the range of the second integer label can be reduced to
$q_c|M_c|$, and the combination $m-t_i$ for $(i,m)$ works as an gauge invariant
quantity. Such $\tilde\cD$ can be viewed as a version of ``semi-direct
product'' of $\cC$ with $\Z_{q_c|M_c|}$. We use the gauge invariant $\tilde
m=m-t_i$ instead of the integer $m$ to label anyons in $\tilde \cD$; in other
words, the anyons are labeled by the new pair $(i,\tilde m)$ where $i\in\cC$
and $\tilde m+t_i \in \Z_{q_c|M_c|}$. Fusion is then given by addition
\begin{align}
  (i,\tilde m)\otimes (j,\tilde n)=\oplus_k N^{ij}_k (k,[\tilde
  m+\tilde n]_{q_c|M_c|}),
\end{align}
where $[\cdots]_{q_c|M_c|}$ denotes the residue modulo $q_c|M_c|$. The $F,R$ matrices in
$\tilde{\cD}$ are given by those in $\cC$ modified by appropriate phase
factors. More precisely, let $F^{i_1i_2i_3}_{i_4}$ 
and $R^{i_1i_2}_{i_3}$ be the $F,R$ matrices in $\cC$; then in $\tilde\cD$ we take
\begin{align}
  &\tilde F^{(i_1,\tilde m_1)(i_2,\tilde m_2)(i_3,\tilde m_3)}_{(i_4,\tilde
  m_4)}
  =F^{i_1i_2i_3}_{i_4}\ee^{
    \frac{\pi\ii}{M_c}\tilde m_1(\tilde m_2+\tilde m_3-[\tilde
    m_2+\tilde m_3]_{q_c|M_c|})},\nonumber\\
    &\tilde R^{(i_1,\tilde m_1)(i_2,\tilde m_2)}_{(i_3,\tilde
    m_3)}=R^{i_1i_2}_{i_3}\ee^{\frac{\pi\ii}{M_c}\tilde
    m_1\tilde m_2}
    =R^{i_1i_2}_{i_3}\ee^{\frac{\pi\ii}{M_c}(m_1-t_{i_1})(m_2-t_{i_2})}.
\end{align}
It is straightforward to check that they satisfy the pentagon and hexagon
equations, and $\tilde \cD$ is a valid pre-modular category. Moreover, the
modified $R$ matrices do give us the desired modified spin. This also suggests
that the hierarchy construction equally works for pre-modular categories, thus
can be easily generalized to topological orders with
symmetries\cite{LW150704673,LW160205936,LW160205946}.

The second step is to reduce the pre-modular category $\tilde\cD$ to the
modular category $\cD$.  Categorically, just note that $\{(i=a_c,\tilde
m=M_c)^{\otimes q},q=0,\dots,q_c-1\}$ forms the M\"{u}ger center of
$\tilde\cD$, which can be identified with $\Rep(\Z_{q_c})$; by condensing this
$\Rep(\Z_{q_c})$ we obtain the desired modular category $\cD$.  Put it simply,
we just further impose the equivalence relation \eqref{equivq} in $\tilde\cD$,
such that one orbit of length $q_c$ is viewed as one type of anyon instead of
$q_c$ different types.  This way we rigorously recover the same construction
described by \eqref{spinim}, \eqref{Sd}, and \eqref{diffc}.

\section{Tables of non-Abelian families}
\label{tables}

In this section, we list some non-Abelian families. 
Each table contains a family up to a certain $N$.
Each row corresponds to a topological order.  The anyons are listed with
increasing quantum dimensions.  Only the quantum dimensions of a root state is
explicitly given. The quantum dimensions of other topological orders can be
easily obtained from those of the root, since the degeneracy for each
value of quantum dimension scales linearly with $N$.
The anyons in red have trivial mutual statistics with all Abelian anyons.
Such sets of anyons are the same within each family, and is an invariant
of the non-Abelian family.

The following is the Abelian family:\\
\centerline{
\def\arraystretch{1.25} \setlength\tabcolsep{3pt}
\begin{tabular}{ |c|c|l|l| }
\hline
$N_{c}$ & $D^2$ & $s_1,s_2,\cdots$ \hfill \blue{$d_1,d_2,\cdots$} \\ 
\hline
$1_{0}$ & $1$ & ${\red {0}}$ \hfill \blue{$1$} \\
$2_{1}$ & $2$ & ${\red {0}},\frac{1}{4}$  \\
$2_{7}$ & $2$ & ${\red {0}},\frac{3}{4}$  \\
$3_{2}$ & $3$ & ${\red {0}},\frac{1}{3},\frac{1}{3}$  \\
$3_{6}$ & $3$ & ${\red {0}},\frac{2}{3},\frac{2}{3}$  \\
$4_{0}$ & $4$ & ${\red {0}},0,0,\frac{1}{2}$  \\
$4_{0}$ & $4$ & ${\red {0}},0,\frac{1}{4},\frac{3}{4}$  \\
$4_{1}$ & $4$ & ${\red {0}},\frac{1}{8},\frac{1}{8},\frac{1}{2}$  \\
$4_{7}$ & $4$ & ${\red {0}},\frac{7}{8},\frac{7}{8},\frac{1}{2}$  \\
$4_{2}$ & $4$ & ${\red {0}},\frac{1}{4},\frac{1}{4},\frac{1}{2}$  \\
$4_{6}$ & $4$ & ${\red {0}},\frac{3}{4},\frac{3}{4},\frac{1}{2}$  \\
$4_{3}$ & $4$ & ${\red {0}},\frac{3}{8},\frac{3}{8},\frac{1}{2}$  \\
$4_{5}$ & $4$ & ${\red {0}},\frac{5}{8},\frac{5}{8},\frac{1}{2}$  \\
$4_{4}$ & $4$ & ${\red {0}},\frac{1}{2},\frac{1}{2},\frac{1}{2}$  \\
$5_{0}$ & $5$ & ${\red {0}},\frac{1}{5},\frac{1}{5},\frac{4}{5},\frac{4}{5}$  \\
$5_{4}$ & $5$ & ${\red {0}},\frac{2}{5},\frac{2}{5},\frac{3}{5},\frac{3}{5}$  \\
$6_{1}$ & $6$ & ${\red {0}},\frac{1}{12},\frac{1}{12},\frac{3}{4},\frac{1}{3},\frac{1}{3}$  \\
$6_{7}$ & $6$ & ${\red {0}},\frac{11}{12},\frac{11}{12},\frac{1}{4},\frac{2}{3},\frac{2}{3}$  \\
$6_{3}$ & $6$ & ${\red {0}},\frac{1}{4},\frac{1}{3},\frac{1}{3},\frac{7}{12},\frac{7}{12}$  \\
$6_{5}$ & $6$ & ${\red {0}},\frac{3}{4},\frac{2}{3},\frac{2}{3},\frac{5}{12},\frac{5}{12}$  \\
$7_{2}$ & $7$ & ${\red {0}},\frac{1}{7},\frac{1}{7},\frac{2}{7},\frac{2}{7},\frac{4}{7},\frac{4}{7}$  \\
$7_{6}$ & $7$ & ${\red {0}},\frac{6}{7},\frac{6}{7},\frac{5}{7},\frac{5}{7},\frac{3}{7},\frac{3}{7}$  \\
$8_{0}$ & $8$ & ${\red {0}},\frac{1}{8},\frac{1}{8},\frac{7}{8},\frac{7}{8},\frac{1}{4},\frac{3}{4},\frac{1}{2}$  \\
$8_{1}$ & $8$ & ${\red {0}},0,0,\frac{1}{4},\frac{1}{4},\frac{1}{4},\frac{3}{4},\frac{1}{2}$  \\
$8_{1}$ & $8$ & ${\red {0}},0,\frac{1}{16},\frac{1}{16},\frac{1}{4},\frac{1}{4},\frac{9}{16},\frac{9}{16}$  \\
$8_{1}$ & $8$ & ${\red {0}},0,\frac{13}{16},\frac{13}{16},\frac{1}{4},\frac{1}{4},\frac{5}{16},\frac{5}{16}$  \\
$8_{7}$ & $8$ & ${\red {0}},0,0,\frac{1}{4},\frac{3}{4},\frac{3}{4},\frac{3}{4},\frac{1}{2}$  \\
$8_{7}$ & $8$ & ${\red {0}},0,\frac{15}{16},\frac{15}{16},\frac{3}{4},\frac{3}{4},\frac{7}{16},\frac{7}{16}$  \\
$8_{7}$ & $8$ & ${\red {0}},0,\frac{3}{16},\frac{3}{16},\frac{3}{4},\frac{3}{4},\frac{11}{16},\frac{11}{16}$  \\
$8_{2}$ & $8$ & ${\red {0}},\frac{1}{8},\frac{1}{8},\frac{1}{4},\frac{3}{4},\frac{3}{8},\frac{3}{8},\frac{1}{2}$  \\
$8_{6}$ & $8$ & ${\red {0}},\frac{7}{8},\frac{7}{8},\frac{1}{4},\frac{3}{4},\frac{5}{8},\frac{5}{8},\frac{1}{2}$  \\
$8_{3}$ & $8$ & ${\red {0}},\frac{1}{4},\frac{1}{4},\frac{1}{4},\frac{3}{4},\frac{1}{2},\frac{1}{2},\frac{1}{2}$  \\
$8_{5}$ & $8$ & ${\red {0}},\frac{1}{4},\frac{3}{4},\frac{3}{4},\frac{3}{4},\frac{1}{2},\frac{1}{2},\frac{1}{2}$  \\
$8_{4}$ & $8$ & ${\red {0}},\frac{1}{4},\frac{3}{4},\frac{3}{8},\frac{3}{8},\frac{5}{8},\frac{5}{8},\frac{1}{2}$  \\
\hline
\end{tabular}
}

~

The following non-Abelian family is described by effective $SU(2)_{-3}$
Chern-Simons (CS) theory plus some Abelian CS theories.  So it is called the
$SU(2)_{-3}$ non-Abelian family.  Due to the level-rank duality, it is also
called the $SU(3)_{2}$ non-Abelian family since its contains a state described
by $SU(3)_{2}$ CS theory. We can also call the family as the Fibonacci
non-Abelian family since the root state is the Fibonacci non-Abelian state.
This family contains FQH state \cite{Wnab,W9927}
\begin{align}
\label{chi32}
\Psi_{\nu=\frac32}(\{z_i\})=[\chi_3(\{z_i\})]^2, \ \ N_c =4_{\frac {21}{5}}
\end{align}

In general, for a state 
\begin{align}
 \Psi_{\nu=\frac{k}{n}}(\{ z_i\})=[\chi_k(\{ z_i\})]^n,
\end{align}
its low energy effective theory obtained from the projective parton
construction is given by \cite{Wnab,W9927}
\begin{align}
\label{chi32L}
 \cL &= \ii 
\psi_{a\al}^\dag (\prt_0 \del_{\al\bt}-\ii (a_0)_{\al\bt} ) \psi_{a\al}
\nonumber\\
&\ \ \ \ 
-\frac{ 
|[(\prt_i -\ii A_i) \del_{\al\bt}-\ii (a_i)_{\al\bt} ]  \psi_{a\bt}|^2}{2m}
\end{align}
where $a=1,\cdots,k=3$ and $\al=1,\cdots,n=2$, and $a_\mu$ is the $SU(n)$ gauge
field doing the projection.  Before the projection (\ie when $a_\mu=0$) the
above effective theory describes a filling fraction $\nu=kn$ IQH state whose
edge has a chiral central charge $c=kn$ (\ie has $kn$ right-moving modes).  After the  projection (\ie after
integrating out the non-zero dynamical $SU(n)$ gauge field $a_\mu$), the edge
states will have a reduced central charge
\begin{align}
 c = kn - \frac{k (n^2-1) }{k+n}.
\end{align}
For our case here, $k=3$ and $n=2$ and we find $c=\frac{21}{5}$ in
\eqref{chi32}.

If we integrate out that fermion fields $\psi_{a\al}$ first, we will obtain an
effective $SU(n)$ CS theory at level $-k$ with central charge $-
\frac{k (n^2-1) }{k+n}$.  The state \eqref{chi32} and the effective theory
\eqref{chi32L} has the same number of anyon types as the $SU(n)_{-k}$ CS theory
But the spins of the anyons in \eqref{chi32}
and in \eqref{chi32L} is not given by those of $SU(n)_{-k}$ CS theory.  They
may differ by $\frac12$ since the anyons in \eqref{chi32L} may contain an extra
fermion field $\psi_{a\al}$.  So the spins of the anyons in \eqref{chi32} and
in \eqref{chi32L} are related to the spins in $SU(n)_{-k}$ CS theory via
\begin{align}
 s_i = s_i^{SU(n)_{-k}} \text{ mod } \frac 12.
\end{align}
In other words, 
the spins of the anyons in \eqref{chi32} and
in \eqref{chi32L} are related to the spins in $SU(n)_{k}$ CS theory via
\begin{align}
 s_i = - s_i^{SU(n)_{k}} \text{ mod } \frac 12.
\end{align}
This allows us to identify the state \eqref{chi32} in the table of the
non-Abelian family which is marked by the red $N_c$.  We will denote the
effective $SU(n)$ CS theory obtained from \eqref{chi32} by integrating out the
fermions as $SU(n)_{-k}^f$.  So the red $N_c=4_\frac{21}{5}$ row in the
following table is described by the $SU(n)_{-k}^f$ CS effective theory
\eqref{chi32L}. On the other hand, the row $N_c=4_\frac{31}{5}$ is described by
the pure $SU(n)_{-k}$ CS effective theory (\ie without coupling to fermionic
fields).

\centerline{
\def\arraystretch{1.25} \setlength\tabcolsep{3pt}
\begin{tabular}{ |c|c|l|l| }
\hline
$N_{c}$ & $D^2$ & $s_1,s_2,\cdots$ \hfill \blue{$d_1,d_2,\cdots$} \\ 
\hline
$2_{\frac{26}{5}}$ & $3.618$ & ${\red {0}},{\red {\frac{3}{5}}}$ \hfill \blue{$1,\zeta_{3}^{1}$} \\
$4_{\frac{31}{5}}$ & $7.236$ & ${\red {0}},\frac{1}{4},\frac{17}{20},{\red {\frac{3}{5}}}$   \hfill $SU(2)_{-3}$ CS\\
$\red 4_{\frac{21}{5}}$ & $7.236$ & ${\red {0}},\frac{3}{4},\frac{7}{20},{\red {\frac{3}{5}}}$   \hfill $SU(2)_{-3}^f$ CS\\
$6_{\frac{36}{5}}$ & $10.85$ & ${\red {0}},\frac{1}{3},\frac{1}{3},\frac{14}{15},\frac{14}{15},{\red {\frac{3}{5}}}$  \\
$6_{\frac{16}{5}}$ & $10.85$ & ${\red {0}},\frac{2}{3},\frac{2}{3},\frac{4}{15},\frac{4}{15},{\red {\frac{3}{5}}}$  \\
$8_{\frac{1}{5}}$ & $14.47$ & ${\red {0}},\frac{3}{8},\frac{3}{8},\frac{1}{2},\frac{39}{40},\frac{39}{40},\frac{1}{10},{\red {\frac{3}{5}}}$  \\
$8_{\frac{36}{5}}$ & $14.47$ & ${\red {0}},\frac{1}{4},\frac{1}{4},\frac{1}{2},\frac{1}{10},\frac{17}{20},\frac{17}{20},{\red {\frac{3}{5}}}$  \\
\hline
\end{tabular}
}

~

The following non-Abelian family is described by effective $SU(2)_3$ CS theory.
So it is called the $SU(2)_3$ non-Abelian family.  The states in the following
table are time-reversal conjugates of those in the previous table.  This family
contains FQH state \cite{Wnab,W9927}
\begin{align}
\Psi_{\nu=-\frac32}(\{\bar z_i\})=[\chi_3(\{\bar z_i\})]^2, \ \
N_c =4_{\frac {19}{5}}
\end{align}

\centerline{
\def\arraystretch{1.25} \setlength\tabcolsep{3pt}
\begin{tabular}{ |c|c|l|l| }
\hline
$N_{c}$ & $D^2$ & $s_1,s_2,\cdots$ \hfill \blue{$d_1,d_2,\cdots$} \\ 
\hline
$2_{\frac{14}{5}}$ & $3.618$ & ${\red {0}},{\red {\frac{2}{5}}}$ \hfill \blue{$1,\zeta_{3}^{1}$} \\
$4_{\frac{9}{5}}$ & $7.236$ & ${\red {0}},\frac{3}{4},\frac{3}{20},{\red {\frac{2}{5}}}$  \hfill $SU(2)_3$ CS\\
${\red 4_{\frac{19}{5}}}$ & $7.236$ & ${\red {0}},\frac{1}{4},\frac{13}{20},{\red {\frac{2}{5}}}$  \hfill $SU(2)_3^f$ CS\\
$6_{\frac{4}{5}}$ & $10.85$ & ${\red {0}},\frac{2}{3},\frac{2}{3},\frac{1}{15},\frac{1}{15},{\red {\frac{2}{5}}}$  \\
$6_{\frac{24}{5}}$ & $10.85$ & ${\red {0}},\frac{1}{3},\frac{1}{3},\frac{11}{15},\frac{11}{15},{\red {\frac{2}{5}}}$  \\
$8_{\frac{39}{5}}$ & $14.47$ & ${\red {0}},\frac{5}{8},\frac{5}{8},\frac{1}{2},\frac{1}{40},\frac{1}{40},\frac{9}{10},{\red {\frac{2}{5}}}$  \\
$8_{\frac{4}{5}}$ & $14.47$ & ${\red {0}},\frac{3}{4},\frac{3}{4},\frac{1}{2},\frac{9}{10},\frac{3}{20},\frac{3}{20},{\red {\frac{2}{5}}}$  \\
\hline
\end{tabular}
}

\vfill\break

The following $SU(2)_2$ Ising non-Abelian family contains FQH states \cite{Wnab,W9927,MR9162}
\begin{align}
\Psi_{\nu=1}(\{z_i\})=[\chi_2(\{z_i\})]^2,\ \ N_c=3_{\frac 52} 
\end{align}
and
\begin{align} 
\Psi_{\nu=1} = \cA(\frac{1}{z_1-z_2} \frac{1}{z_2-z_3} \cdots)\prod (z_i-z_j),
,\ \ N_c=3_{\frac 32} 
\end{align} 
\centerline{
\def\arraystretch{1.25} \setlength\tabcolsep{3pt}
\begin{tabular}{ |c|c|l|l| }
\hline
$N_{c}$ & $D^2$ & $s_1,s_2,\cdots$ \hfill \blue{$d_1,d_2,\cdots$} \\ 
\hline
$3_{\frac{1}{2}}$ & $4$ & ${\red {0}},{\red {\frac{1}{2}}},\frac{1}{16}$ \hfill \blue{$1,1,\zeta_{2}^{1}$} \\
$3_{\frac{15}{2}}$ & $4$ & ${\red {0}},{\red {\frac{1}{2}}},\frac{15}{16}$  \\
$\red 3_{\frac{3}{2}}$ & $4$ & ${\red {0}},{\red {\frac{1}{2}}},\frac{3}{16}$  \\
$3_{\frac{13}{2}}$ & $4$ & ${\red {0}},{\red {\frac{1}{2}}},\frac{13}{16}$  \\
$\red 3_{\frac{5}{2}}$ & $4$ & ${\red {0}},{\red {\frac{1}{2}}},\frac{5}{16}$  \\
$3_{\frac{11}{2}}$ & $4$ & ${\red {0}},{\red {\frac{1}{2}}},\frac{11}{16}$  \\
$3_{\frac{7}{2}}$ & $4$ & ${\red {0}},{\red {\frac{1}{2}}},\frac{7}{16}$  \\
\hline
\end{tabular}
}

\vfill
\break

The following $SU(2)_5$ non-Abelian family contains FQH state \cite{Wnab,W9927}
\begin{align}
\Psi_{\nu=-\frac 52}(\{\bar z_i\})=[\chi_5(\{\bar z_i\})]^2,\ \ N_c=6_{\frac {1}{7}} 
\end{align}
\centerline{
\def\arraystretch{1.25} \setlength\tabcolsep{3pt}
\begin{tabular}{ |c|c|l|l| }
\hline
$N_{c}$ & $D^2$ & $s_1,s_2,\cdots$ \hfill \blue{$d_1,d_2,\cdots$} \\ 
\hline
$3_{\frac{8}{7}}$ & $9.295$ & ${\red {0}},{\red {\frac{6}{7}}},{\red {\frac{2}{7}}}$ \hfill \blue{$1,\zeta_{5}^{1},\zeta_{5}^{2}$} \\
$\red 6_{\frac{1}{7}}$ & $18.59$ & ${\red {0}},\frac{3}{4},{\red {\frac{6}{7}}},\frac{17}{28},\frac{1}{28},{\red {\frac{2}{7}}}$  \\
$6_{\frac{15}{7}}$ & $18.59$ & ${\red {0}},\frac{1}{4},\frac{3}{28},{\red {\frac{6}{7}}},{\red {\frac{2}{7}}},\frac{15}{28}$  \\
$9_{\frac{50}{7}}$ & $27.88$ & ${\red {0}},\frac{2}{3},\frac{2}{3},{\red {\frac{6}{7}}},\frac{11}{21},\frac{11}{21},\frac{20}{21},\frac{20}{21},{\red {\frac{2}{7}}}$  \\
$9_{\frac{22}{7}}$ & $27.88$ & ${\red {0}},\frac{1}{3},\frac{1}{3},{\red {\frac{6}{7}}},\frac{4}{21},\frac{4}{21},{\red {\frac{2}{7}}},\frac{13}{21},\frac{13}{21}$  \\
$12_{\frac{1}{7}}$ & $37.18$ & ${\red {0}},\frac{7}{8},\frac{7}{8},\frac{1}{2},{\red {\frac{6}{7}}},\frac{41}{56},\frac{41}{56},\frac{5}{14},\frac{9}{56},\frac{9}{56},\frac{11}{14},{\red {\frac{2}{7}}}$  \\
$12_{\frac{50}{7}}$ & $37.18$ & ${\red {0}},\frac{3}{4},\frac{3}{4},\frac{1}{2},{\red {\frac{6}{7}}},\frac{5}{14},\frac{17}{28},\frac{17}{28},\frac{1}{28},\frac{1}{28},\frac{11}{14},{\red {\frac{2}{7}}}$  \\
\hline
\end{tabular}
}

~

~

The following $SU(2)_{-5}$ non-Abelian family (or $SU(5)_{2}$ non-Abelian family) contains FQH state \cite{Wnab,W9927}
\begin{align}
\Psi_{\nu=\frac 52}(\{z_i\})=[\chi_5(\{z_i\})]^2,\ \ N_c=6_{\frac {55}{7}} 
\end{align}
\centerline{
\def\arraystretch{1.25} \setlength\tabcolsep{3pt}
\begin{tabular}{ |c|c|l|l| }
\hline
$N_{c}$ & $D^2$ & $s_1,s_2,\cdots$ \hfill \blue{$d_1,d_2,\cdots$} \\ 
\hline
$3_{\frac{48}{7}}$ & $9.295$ & ${\red {0}},{\red {\frac{1}{7}}},{\red {\frac{5}{7}}}$ \hfill \blue{$1,\zeta_{5}^{1},\zeta_{5}^{2}$} \\
$\red 6_{\frac{55}{7}}$ & $18.59$ & ${\red {0}},\frac{1}{4},{\red {\frac{1}{7}}},\frac{11}{28},\frac{27}{28},{\red {\frac{5}{7}}}$  \\
$6_{\frac{41}{7}}$ & $18.59$ & ${\red {0}},\frac{3}{4},\frac{25}{28},{\red {\frac{1}{7}}},{\red {\frac{5}{7}}},\frac{13}{28}$  \\
$9_{\frac{6}{7}}$ & $27.88$ & ${\red {0}},\frac{1}{3},\frac{1}{3},{\red {\frac{1}{7}}},\frac{10}{21},\frac{10}{21},\frac{1}{21},\frac{1}{21},{\red {\frac{5}{7}}}$  \\
$9_{\frac{34}{7}}$ & $27.88$ & ${\red {0}},\frac{2}{3},\frac{2}{3},{\red {\frac{1}{7}}},\frac{17}{21},\frac{17}{21},{\red {\frac{5}{7}}},\frac{8}{21},\frac{8}{21}$  \\
$12_{\frac{55}{7}}$ & $37.18$ & ${\red {0}},\frac{1}{8},\frac{1}{8},\frac{1}{2},{\red {\frac{1}{7}}},\frac{15}{56},\frac{15}{56},\frac{9}{14},\frac{47}{56},\frac{47}{56},\frac{3}{14},{\red {\frac{5}{7}}}$  \\
$12_{\frac{6}{7}}$ & $37.18$ & ${\red {0}},\frac{1}{4},\frac{1}{4},\frac{1}{2},{\red {\frac{1}{7}}},\frac{9}{14},\frac{11}{28},\frac{11}{28},\frac{27}{28},\frac{27}{28},\frac{3}{14},{\red {\frac{5}{7}}}$  \\
\hline
\end{tabular}
}

\vfill
\break

The following three  non-Abelian families are obtained
by stacking two FQH states from the two 
Fibonacci non-Abelian families.\\
\centerline{
\def\arraystretch{1.25} \setlength\tabcolsep{3pt}
\begin{tabular}{ |c|c|l|l| }
\hline
$N_{c}$ & $D^2$ & $s_1,s_2,\cdots$ \hfill \blue{$d_1,d_2,\cdots$} \\ 
\hline
$4_{0}$ & $13.09$ & ${\red {0}},{\red {\frac{2}{5}}},{\red {\frac{3}{5}}},{\red {0}}$ \hfill \blue{$1,\zeta_{3}^{1},\zeta_{3}^{1},\zeta_{8}^{2}$} \\
$8_{1}$ & $26.18$ & ${\red {0}},\frac{1}{4},\frac{17}{20},\frac{13}{20},{\red {\frac{2}{5}}},{\red {\frac{3}{5}}},{\red {0}},\frac{1}{4}$  \\
$8_{7}$ & $26.18$ & ${\red {0}},\frac{3}{4},\frac{3}{20},\frac{7}{20},{\red {\frac{2}{5}}},{\red {\frac{3}{5}}},{\red {0}},\frac{3}{4}$  \\
$12_{2}$ & $39.27$ & ${\red {0}},\frac{1}{3},\frac{1}{3},\frac{14}{15},\frac{14}{15},\frac{11}{15},\frac{11}{15},{\red {\frac{2}{5}}},{\red {\frac{3}{5}}},{\red {0}},\frac{1}{3},\frac{1}{3}$  \\
$12_{6}$ & $39.27$ & ${\red {0}},\frac{2}{3},\frac{2}{3},\frac{1}{15},\frac{1}{15},\frac{4}{15},\frac{4}{15},{\red {\frac{2}{5}}},{\red {\frac{3}{5}}},{\red {0}},\frac{2}{3},\frac{2}{3}$  \\
\hline
\end{tabular}
}

~

\centerline{
\def\arraystretch{1.25} \setlength\tabcolsep{3pt}
\begin{tabular}{ |c|c|l|l| }
\hline
$N_{c}$ & $D^2$ & $s_1,s_2,\cdots$ \hfill \blue{$d_1,d_2,\cdots$} \\ 
\hline
$4_{\frac{12}{5}}$ & $13.09$ & ${\red {0}},{\red {\frac{3}{5}}},{\red {\frac{3}{5}}},{\red {\frac{1}{5}}}$ \hfill \blue{$1,\zeta_{3}^{1},\zeta_{3}^{1},\zeta_{8}^{2}$} \\
$8_{\frac{7}{5}}$ & $26.18$ & ${\red {0}},\frac{3}{4},\frac{7}{20},\frac{7}{20},{\red {\frac{3}{5}}},{\red {\frac{3}{5}}},\frac{19}{20},{\red {\frac{1}{5}}}$  \\
$8_{\frac{17}{5}}$ & $26.18$ & ${\red {0}},\frac{1}{4},\frac{17}{20},\frac{17}{20},{\red {\frac{3}{5}}},{\red {\frac{3}{5}}},{\red {\frac{1}{5}}},\frac{9}{20}$  \\
$12_{\frac{2}{5}}$ & $39.27$ & ${\red {0}},\frac{2}{3},\frac{2}{3},\frac{4}{15},\frac{4}{15},\frac{4}{15},\frac{4}{15},{\red {\frac{3}{5}}},{\red {\frac{3}{5}}},\frac{13}{15},\frac{13}{15},{\red {\frac{1}{5}}}$  \\
$12_{\frac{22}{5}}$ & $39.27$ & ${\red {0}},\frac{1}{3},\frac{1}{3},\frac{14}{15},\frac{14}{15},\frac{14}{15},\frac{14}{15},{\red {\frac{3}{5}}},{\red {\frac{3}{5}}},{\red {\frac{1}{5}}},\frac{8}{15},\frac{8}{15}$  \\
\hline
\end{tabular}
}

~

\centerline{
\def\arraystretch{1.25} \setlength\tabcolsep{3pt}
\begin{tabular}{ |c|c|l|l| }
\hline
$N_{c}$ & $D^2$ & $s_1,s_2,\cdots$ \hfill \blue{$d_1,d_2,\cdots$} \\ 
\hline
$4_{\frac{28}{5}}$ & $13.09$ & ${\red {0}},{\red {\frac{2}{5}}},{\red {\frac{2}{5}}},{\red {\frac{4}{5}}}$ \hfill \blue{$1,\zeta_{3}^{1},\zeta_{3}^{1},\zeta_{8}^{2}$} \\
$8_{\frac{33}{5}}$ & $26.18$ & ${\red {0}},\frac{1}{4},\frac{13}{20},\frac{13}{20},{\red {\frac{2}{5}}},{\red {\frac{2}{5}}},\frac{1}{20},{\red {\frac{4}{5}}}$  \\
$8_{\frac{23}{5}}$ & $26.18$ & ${\red {0}},\frac{3}{4},\frac{3}{20},\frac{3}{20},{\red {\frac{2}{5}}},{\red {\frac{2}{5}}},{\red {\frac{4}{5}}},\frac{11}{20}$  \\
$12_{\frac{38}{5}}$ & $39.27$ & ${\red {0}},\frac{1}{3},\frac{1}{3},\frac{11}{15},\frac{11}{15},\frac{11}{15},\frac{11}{15},{\red {\frac{2}{5}}},{\red {\frac{2}{5}}},\frac{2}{15},\frac{2}{15},{\red {\frac{4}{5}}}$  \\
$12_{\frac{18}{5}}$ & $39.27$ & ${\red {0}},\frac{2}{3},\frac{2}{3},\frac{1}{15},\frac{1}{15},\frac{1}{15},\frac{1}{15},{\red {\frac{2}{5}}},{\red {\frac{2}{5}}},{\red {\frac{4}{5}}},\frac{7}{15},\frac{7}{15}$  \\
\hline
\end{tabular}
}

~

~

The following $SU(2)_7$ non-Abelian family contains FQH state \cite{Wnab,W9927}
\begin{align}
\Psi_{\nu=-\frac 72}(\{\bar z_i\})=[\chi_7(\{\bar z_i\})]^2,\ \ N_c=8_{\frac {13}{3}} 
\end{align}
\centerline{
\def\arraystretch{1.25} \setlength\tabcolsep{3pt}
\begin{tabular}{ |c|c|l|l| }
\hline
$N_{c}$ & $D^2$ & $s_1,s_2,\cdots$ \hfill \blue{$d_1,d_2,\cdots$} \\ 
\hline
$4_{\frac{10}{3}}$ & $19.23$ & ${\red {0}},{\red {\frac{1}{3}}},{\red {\frac{2}{9}}},{\red {\frac{2}{3}}}$ \hfill \blue{$1,\zeta_{7}^{1},\zeta_{7}^{2},\zeta_{7}^{3}$} \\
$8_{\frac{7}{3}}$ & $38.46$ & ${\red {0}},\frac{3}{4},\frac{1}{12},{\red {\frac{1}{3}}},\frac{35}{36},{\red {\frac{2}{9}}},{\red {\frac{2}{3}}},\frac{5}{12}$  \\
$\red 8_{\frac{13}{3}}$ & $38.46$ & ${\red {0}},\frac{1}{4},{\red {\frac{1}{3}}},\frac{7}{12},{\red {\frac{2}{9}}},\frac{17}{36},\frac{11}{12},{\red {\frac{2}{3}}}$  \\
$12_{\frac{4}{3}}$ & $57.70$ & ${\red {0}},\frac{2}{3},\frac{2}{3},0,0,{\red {\frac{1}{3}}},\frac{8}{9},\frac{8}{9},{\red {\frac{2}{9}}},\frac{1}{3},\frac{1}{3},{\red {\frac{2}{3}}}$  \\
$12_{\frac{16}{3}}$ & $57.70$ & ${\red {0}},\frac{1}{3},\frac{1}{3},{\red {\frac{1}{3}}},\frac{2}{3},\frac{2}{3},{\red {\frac{2}{9}}},\frac{5}{9},\frac{5}{9},0,0,{\red {\frac{2}{3}}}$  \\
\hline
\end{tabular}
}

~

~

The following $SU(2)_{-7}$ non-Abelian family contains FQH state \cite{Wnab,W9927}
\begin{align}
\Psi_{\nu=\frac 72}(\{z_i\})=[\chi_7(\{z_i\})]^2,\ \ N_c=8_{\frac {35}{3}} 
\end{align}
\centerline{
\def\arraystretch{1.25} \setlength\tabcolsep{3pt}
\begin{tabular}{ |c|c|l|l| }
\hline
$N_{c}$ & $D^2$ & $s_1,s_2,\cdots$ \hfill \blue{$d_1,d_2,\cdots$} \\ 
\hline
$4_{\frac{14}{3}}$ & $19.23$ & ${\red {0}},{\red {\frac{2}{3}}},{\red {\frac{7}{9}}},{\red {\frac{1}{3}}}$ \hfill \blue{$1,\zeta_{7}^{1},\zeta_{7}^{2},\zeta_{7}^{3}$} \\
$8_{\frac{17}{3}}$ & $38.46$ & ${\red {0}},\frac{1}{4},\frac{11}{12},{\red {\frac{2}{3}}},\frac{1}{36},{\red {\frac{7}{9}}},{\red {\frac{1}{3}}},\frac{7}{12}$  \\
$\red 8_{\frac{11}{3}}$ & $38.46$ & ${\red {0}},\frac{3}{4},{\red {\frac{2}{3}}},\frac{5}{12},{\red {\frac{7}{9}}},\frac{19}{36},\frac{1}{12},{\red {\frac{1}{3}}}$  \\
$12_{\frac{20}{3}}$ & $57.70$ & ${\red {0}},\frac{1}{3},\frac{1}{3},0,0,{\red {\frac{2}{3}}},\frac{1}{9},\frac{1}{9},{\red {\frac{7}{9}}},{\red {\frac{1}{3}}},\frac{2}{3},\frac{2}{3}$  \\
$12_{\frac{8}{3}}$ & $57.70$ & ${\red {0}},\frac{2}{3},\frac{2}{3},\frac{1}{3},\frac{1}{3},{\red {\frac{2}{3}}},{\red {\frac{7}{9}}},\frac{4}{9},\frac{4}{9},0,0,{\red {\frac{1}{3}}}$  \\
\hline
\end{tabular}
}

\vfill
\break

The following $SU(2)_4$ non-Abelian family contains FQH state \cite{Wnab,W9927}
\begin{align}
\Psi_{\nu=\frac 12}(\{z_i\})=[\chi_2(\{z_i\})]^4,\ \ N_c=10_{3} 
\end{align}
\centerline{
\def\arraystretch{1.25} \setlength\tabcolsep{3pt}
\begin{tabular}{ |c|c|l|l| }
\hline
$N_{c}$ & $D^2$ & $s_1,s_2,\cdots$ \hfill \blue{$d_1,d_2,\cdots$} \\ 
\hline
$5_{2}$ & $12$ & ${\red {0}},{\red {0}},\frac{1}{8},\frac{5}{8},{\red {\frac{1}{3}}}$ \hfill \blue{$1,1,\zeta_{4}^{1},\zeta_{4}^{1},2$} \\
$5_{2}$ & $12$ & ${\red {0}},{\red {0}},\frac{7}{8},\frac{3}{8},{\red {\frac{1}{3}}}$  \\
$10_{1}$ & $24$ & ${\red {0}},{\red {0}},\frac{3}{4},\frac{3}{4},\frac{1}{8},\frac{7}{8},\frac{3}{8},\frac{5}{8},\frac{1}{12},{\red {\frac{1}{3}}}$  \\
$10_{1}$ & $24$ & ${\red {0}},{\red {0}},\frac{3}{4},\frac{3}{4},\frac{1}{16},\frac{1}{16},\frac{9}{16},\frac{9}{16},\frac{1}{12},{\red {\frac{1}{3}}}$  \\
$10_{1}$ & $24$ & ${\red {0}},{\red {0}},\frac{3}{4},\frac{3}{4},\frac{13}{16},\frac{13}{16},\frac{5}{16},\frac{5}{16},\frac{1}{12},{\red {\frac{1}{3}}}$  \\
$10_{3}$ & $24$ & ${\red {0}},{\red {0}},\frac{1}{4},\frac{1}{4},\frac{1}{8},\frac{7}{8},\frac{3}{8},\frac{5}{8},{\red {\frac{1}{3}}},\frac{7}{12}$  \\
$\red 10_{3}$ & $24$ & ${\red {0}},{\red {0}},\frac{1}{4},\frac{1}{4},\frac{3}{16},\frac{3}{16},\frac{11}{16},\frac{11}{16},{\red {\frac{1}{3}}},\frac{7}{12}$  \hfill $SU(4)_{-2}^f$, $SU(4)_{-2}$ \\
\hline
\end{tabular}
}

~

~

The following $SU(2)_{-4}$ non-Abelian family contains FQH state \cite{Wnab,W9927}
\begin{align}
\Psi_{\nu=-\frac 12}(\{\bar z_i\})=[\chi_2(\{\bar z_i\})]^4,\ \ N_c=10_{5} 
\end{align}
\centerline{
\def\arraystretch{1.25} \setlength\tabcolsep{3pt}
\begin{tabular}{ |c|c|l|l| }
\hline
$N_{c}$ & $D^2$ & $s_1,s_2,\cdots$ \hfill \blue{$d_1,d_2,\cdots$} \\ 
\hline
$5_{6}$ & $12$ & ${\red {0}},{\red {0}},\frac{7}{8},\frac{3}{8},{\red {\frac{2}{3}}}$ \hfill \blue{$1,1,\zeta_{4}^{1},\zeta_{4}^{1},2$} \\
$5_{6}$ & $12$ & ${\red {0}},{\red {0}},\frac{1}{8},\frac{5}{8},{\red {\frac{2}{3}}}$  \\
$10_{7}$ & $24$ & ${\red {0}},{\red {0}},\frac{1}{4},\frac{1}{4},\frac{1}{8},\frac{7}{8},\frac{3}{8},\frac{5}{8},\frac{11}{12},{\red {\frac{2}{3}}}$  \\
$10_{7}$ & $24$ & ${\red {0}},{\red {0}},\frac{1}{4},\frac{1}{4},\frac{15}{16},\frac{15}{16},\frac{7}{16},\frac{7}{16},\frac{11}{12},{\red {\frac{2}{3}}}$  \\
$10_{7}$ & $24$ & ${\red {0}},{\red {0}},\frac{1}{4},\frac{1}{4},\frac{3}{16},\frac{3}{16},\frac{11}{16},\frac{11}{16},\frac{11}{12},{\red {\frac{2}{3}}}$  \\
$10_{5}$ & $24$ & ${\red {0}},{\red {0}},\frac{3}{4},\frac{3}{4},\frac{1}{8},\frac{7}{8},\frac{3}{8},\frac{5}{8},{\red {\frac{2}{3}}},\frac{5}{12}$  \\
$\red 10_{5}$ & $24$ & ${\red {0}},{\red {0}},\frac{3}{4},\frac{3}{4},\frac{13}{16},\frac{13}{16},\frac{5}{16},\frac{5}{16},{\red {\frac{2}{3}}},\frac{5}{12}$  \hfill $SU(4)_{2}^f$, $SU(4)_2$ \\
\hline
\end{tabular}
}

~

~

The following $SU(2)_9$ non-Abelian family contains FQH state \cite{Wnab,W9927}
\begin{align}
\Psi_{\nu=-\frac 92}(\{\bar z_i\})=[\chi_9(\{\bar z_i\})]^2,\ \ N_c=10_{\frac {5}{11}} 
\end{align}
\centerline{
\def\arraystretch{1.25} \setlength\tabcolsep{3pt}
\begin{tabular}{ |c|c|l|l| }
\hline
$N_{c}$ & $D^2$ & $s_1,s_2,\cdots$ \hfill \blue{$d_1,d_2,\cdots$} \\ 
\hline
$5_{\frac{16}{11}}$ & $34.64$ & ${\red {0}},{\red {\frac{9}{11}}},{\red {\frac{2}{11}}},{\red {\frac{1}{11}}},{\red {\frac{6}{11}}}$ \hfill \blue{$1,\zeta_{9}^{1},\zeta_{9}^{2},\zeta_{9}^{3},\zeta_{9}^{4}$} \\
$\red 10_{\frac{5}{11}}$ & $69.29$ & ${\red {0}},\frac{3}{4},{\red {\frac{9}{11}}},\frac{25}{44},\frac{41}{44},{\red {\frac{2}{11}}},{\red {\frac{1}{11}}},\frac{37}{44},\frac{13}{44},{\red {\frac{6}{11}}}$  \\
$10_{\frac{27}{11}}$ & $69.29$ & ${\red {0}},\frac{1}{4},\frac{3}{44},{\red {\frac{9}{11}}},{\red {\frac{2}{11}}},\frac{19}{44},{\red {\frac{1}{11}}},\frac{15}{44},\frac{35}{44},{\red {\frac{6}{11}}}$  \\
$15_{\frac{82}{11}}$ & $103.9$ & ${\red {0}},\frac{2}{3},\frac{2}{3},{\red {\frac{9}{11}}},\frac{16}{33},\frac{16}{33},\frac{28}{33},\frac{28}{33},{\red {\frac{2}{11}}},{\red {\frac{1}{11}}},\frac{25}{33},\frac{25}{33},\frac{7}{33},\frac{7}{33},{\red {\frac{6}{11}}}$  \\
$15_{\frac{38}{11}}$ & $103.9$ & ${\red {0}},\frac{1}{3},\frac{1}{3},\frac{5}{33},\frac{5}{33},{\red {\frac{9}{11}}},{\red {\frac{2}{11}}},\frac{17}{33},\frac{17}{33},{\red {\frac{1}{11}}},\frac{14}{33},\frac{14}{33},\frac{29}{33},\frac{29}{33},{\red {\frac{6}{11}}}$  \\
\hline
\end{tabular}
}

\vfill
\break

The following $SU(2)_{-9}$ non-Abelian family contains FQH state \cite{Wnab,W9927}
\begin{align}
\Psi_{\nu=\frac 92}(\{z_i\})=[\chi_9(\{z_i\})]^2,\ \ N_c=10_{\frac {83}{11}} 
\end{align}
\centerline{
\def\arraystretch{1.25} \setlength\tabcolsep{3pt}
\begin{tabular}{ |c|c|l|l| }
\hline
$N_{c}$ & $D^2$ & $s_1,s_2,\cdots$ \hfill \blue{$d_1,d_2,\cdots$} \\ 
\hline
$5_{\frac{72}{11}}$ & $34.64$ & ${\red {0}},{\red {\frac{2}{11}}},{\red {\frac{9}{11}}},{\red {\frac{10}{11}}},{\red {\frac{5}{11}}}$ \hfill \blue{$1,\zeta_{9}^{1},\zeta_{9}^{2},\zeta_{9}^{3},\zeta_{9}^{4}$} \\
$\red 10_{\frac{83}{11}}$ & $69.29$ & ${\red {0}},\frac{1}{4},{\red {\frac{2}{11}}},\frac{19}{44},\frac{3}{44},{\red {\frac{9}{11}}},{\red {\frac{10}{11}}},\frac{7}{44},\frac{31}{44},{\red {\frac{5}{11}}}$  \\
$10_{\frac{61}{11}}$ & $69.29$ & ${\red {0}},\frac{3}{4},\frac{41}{44},{\red {\frac{2}{11}}},{\red {\frac{9}{11}}},\frac{25}{44},{\red {\frac{10}{11}}},\frac{29}{44},\frac{9}{44},{\red {\frac{5}{11}}}$  \\
$15_{\frac{6}{11}}$ & $103.9$ & ${\red {0}},\frac{1}{3},\frac{1}{3},{\red {\frac{2}{11}}},\frac{17}{33},\frac{17}{33},\frac{5}{33},\frac{5}{33},{\red {\frac{9}{11}}},{\red {\frac{10}{11}}},\frac{8}{33},\frac{8}{33},\frac{26}{33},\frac{26}{33},{\red {\frac{5}{11}}}$  \\
$15_{\frac{50}{11}}$ & $103.9$ & ${\red {0}},\frac{2}{3},\frac{2}{3},\frac{28}{33},\frac{28}{33},{\red {\frac{2}{11}}},{\red {\frac{9}{11}}},\frac{16}{33},\frac{16}{33},{\red {\frac{10}{11}}},\frac{19}{33},\frac{19}{33},\frac{4}{33},\frac{4}{33},{\red {\frac{5}{11}}}$  \\
\hline
\end{tabular}
}

~

~

The following $SU(3)_{4}$ non-Abelian family contains FQH state \cite{Wnab,W9927}
\begin{align}
\Psi_{\nu=-\frac 74}(\{\bar z_i\})=\chi_1(\{\bar z_i\})[\chi_4(\{\bar z_i\})]^3,\ \ N_c=15_{\frac {4}{7}} 
\end{align}
\centerline{
\def\arraystretch{1.25} \setlength\tabcolsep{3pt}
\begin{tabular}{ |c|c|l|l| }
\hline
$N_{c}$ & $D^2$ & $s_1,s_2,\cdots$ \hfill \blue{$d_1,d_2,\cdots$} \\ 
\hline
$5_{\frac{18}{7}}$ & $35.34$ & ${\red {0}},{\red {\frac{6}{7}}},{\red {\frac{6}{7}}},{\red {\frac{1}{7}}},{\red {\frac{3}{7}}}$ \hfill \blue{$1,\zeta_{5}^{2},\zeta_{5}^{2},\zeta_{12}^{2},\zeta_{12}^{4}$} \\
$10_{\frac{11}{7}}$ & $70.68$ & ${\red {0}},\frac{3}{4},{\red {\frac{6}{7}}},{\red {\frac{6}{7}}},\frac{17}{28},\frac{17}{28},\frac{25}{28},{\red {\frac{1}{7}}},\frac{5}{28},{\red {\frac{3}{7}}}$  \\
$10_{\frac{25}{7}}$ & $70.68$ & ${\red {0}},\frac{1}{4},\frac{3}{28},\frac{3}{28},{\red {\frac{6}{7}}},{\red {\frac{6}{7}}},{\red {\frac{1}{7}}},\frac{11}{28},\frac{19}{28},{\red {\frac{3}{7}}}$  \\
$\red 15_{\frac{4}{7}}$ & $106.0$ & ${\red {0}},\frac{2}{3},\frac{2}{3},{\red {\frac{6}{7}}},{\red {\frac{6}{7}}},\frac{11}{21},\frac{11}{21},\frac{11}{21},\frac{11}{21},{\red {\frac{1}{7}}},\frac{17}{21},\frac{17}{21},\frac{2}{21},\frac{2}{21},{\red {\frac{3}{7}}}$  \\
$15_{\frac{32}{7}}$ & $106.0$ & ${\red {0}},\frac{1}{3},\frac{1}{3},{\red {\frac{6}{7}}},{\red {\frac{6}{7}}},\frac{4}{21},\frac{4}{21},\frac{4}{21},\frac{4}{21},{\red {\frac{1}{7}}},\frac{10}{21},\frac{10}{21},\frac{16}{21},\frac{16}{21},{\red {\frac{3}{7}}}$  \\
\hline
\end{tabular}
}

~

~

The following $SU(3)_{-4}$ non-Abelian family contains FQH state \cite{Wnab,W9927}
\begin{align}
\Psi_{\nu=\frac 74}(\{z_i\})=\chi_1(\{z_i\})[\chi_4(\{z_i\})]^3,\ \ N_c=15_{\frac {52}{7}} 
\end{align}
\centerline{
\def\arraystretch{1.25} \setlength\tabcolsep{3pt}
\begin{tabular}{ |c|c|l|l| }
\hline
$N_{c}$ & $D^2$ & $s_1,s_2,\cdots$ \hfill \blue{$d_1,d_2,\cdots$} \\ 
\hline
$5_{\frac{38}{7}}$ & $35.34$ & ${\red {0}},{\red {\frac{1}{7}}},{\red {\frac{1}{7}}},{\red {\frac{6}{7}}},{\red {\frac{4}{7}}}$ \hfill \blue{$1,\zeta_{5}^{2},\zeta_{5}^{2},\zeta_{12}^{2},\zeta_{12}^{4}$} \\
$10_{\frac{45}{7}}$ & $70.68$ & ${\red {0}},\frac{1}{4},{\red {\frac{1}{7}}},{\red {\frac{1}{7}}},\frac{11}{28},\frac{11}{28},\frac{3}{28},{\red {\frac{6}{7}}},\frac{23}{28},{\red {\frac{4}{7}}}$  \\
$10_{\frac{31}{7}}$ & $70.68$ & ${\red {0}},\frac{3}{4},\frac{25}{28},\frac{25}{28},{\red {\frac{1}{7}}},{\red {\frac{1}{7}}},{\red {\frac{6}{7}}},\frac{17}{28},\frac{9}{28},{\red {\frac{4}{7}}}$  \\
$\red 15_{\frac{52}{7}}$ & $106.0$ & ${\red {0}},\frac{1}{3},\frac{1}{3},{\red {\frac{1}{7}}},{\red {\frac{1}{7}}},\frac{10}{21},\frac{10}{21},\frac{10}{21},\frac{10}{21},{\red {\frac{6}{7}}},\frac{4}{21},\frac{4}{21},\frac{19}{21},\frac{19}{21},{\red {\frac{4}{7}}}$  \\
$15_{\frac{24}{7}}$ & $106.0$ & ${\red {0}},\frac{2}{3},\frac{2}{3},{\red {\frac{1}{7}}},{\red {\frac{1}{7}}},\frac{17}{21},\frac{17}{21},\frac{17}{21},\frac{17}{21},{\red {\frac{6}{7}}},\frac{11}{21},\frac{11}{21},\frac{5}{21},\frac{5}{21},{\red {\frac{4}{7}}}$  \\
\hline
\end{tabular}
}

\begin{widetext}

~

~

~

The following four non-Abelian families are obtained by stacking a FQH state
from the two $SU(2)_{\pm 3}$ non-Abelian families with a FQH state from the two
$SU(2)_{\pm 5}$ non-Abelian families.\\ 
\centerline{
\def\arraystretch{1.25}
\setlength\tabcolsep{3pt}
\begin{tabular}{ |c|c|l|l| }
\hline
$N_{c}$ & $D^2$ & $s_1,s_2,\cdots$ \hfill \blue{$d_1,d_2,\cdots$} \\ 
\hline
$6_{\frac{58}{35}}$ & $33.63$ & ${\red {0}},{\red {\frac{2}{5}}},{\red {\frac{1}{7}}},{\red {\frac{5}{7}}},{\red {\frac{19}{35}}},{\red {\frac{4}{35}}}$ \hfill \blue{$1,\zeta_{3}^{1},\zeta_{5}^{1},\zeta_{5}^{2},2.915,3.635$} \\
$12_{\frac{23}{35}}$ & $67.26$ & ${\red {0}},\frac{3}{4},\frac{3}{20},{\red {\frac{2}{5}}},\frac{25}{28},{\red {\frac{1}{7}}},{\red {\frac{5}{7}}},\frac{13}{28},\frac{41}{140},{\red {\frac{19}{35}}},{\red {\frac{4}{35}}},\frac{121}{140}$  \\
$12_{\frac{93}{35}}$ & $67.26$ & ${\red {0}},\frac{1}{4},\frac{13}{20},{\red {\frac{2}{5}}},{\red {\frac{1}{7}}},\frac{11}{28},\frac{27}{28},{\red {\frac{5}{7}}},\frac{111}{140},{\red {\frac{19}{35}}},{\red {\frac{4}{35}}},\frac{51}{140}$  \\
$18_{\frac{268}{35}}$ & $100.8$ & ${\red {0}},\frac{2}{3},\frac{2}{3},\frac{1}{15},\frac{1}{15},{\red {\frac{2}{5}}},{\red {\frac{1}{7}}},\frac{17}{21},\frac{17}{21},{\red {\frac{5}{7}}},\frac{8}{21},\frac{8}{21},\frac{22}{105},\frac{22}{105},{\red {\frac{19}{35}}},{\red {\frac{4}{35}}},\frac{82}{105},\frac{82}{105}$  \\
$18_{\frac{128}{35}}$ & $100.8$ & ${\red {0}},\frac{1}{3},\frac{1}{3},\frac{11}{15},\frac{11}{15},{\red {\frac{2}{5}}},{\red {\frac{1}{7}}},\frac{10}{21},\frac{10}{21},\frac{1}{21},\frac{1}{21},{\red {\frac{5}{7}}},\frac{92}{105},\frac{92}{105},{\red {\frac{19}{35}}},{\red {\frac{4}{35}}},\frac{47}{105},\frac{47}{105}$  \\
\hline
\end{tabular}
}

~

\centerline{
\def\arraystretch{1.25} \setlength\tabcolsep{3pt}
\begin{tabular}{ |c|c|l|l| }
\hline
$N_{c}$ & $D^2$ & $s_1,s_2,\cdots$ \hfill \blue{$d_1,d_2,\cdots$} \\ 
\hline
$6_{\frac{138}{35}}$ & $33.63$ & ${\red {0}},{\red {\frac{2}{5}}},{\red {\frac{6}{7}}},{\red {\frac{2}{7}}},{\red {\frac{9}{35}}},{\red {\frac{24}{35}}}$ \hfill \blue{$1,\zeta_{3}^{1},\zeta_{5}^{1},\zeta_{5}^{2},2.915,3.635$} \\
$12_{\frac{103}{35}}$ & $67.26$ & ${\red {0}},\frac{3}{4},\frac{3}{20},{\red {\frac{2}{5}}},{\red {\frac{6}{7}}},\frac{17}{28},\frac{1}{28},{\red {\frac{2}{7}}},\frac{1}{140},{\red {\frac{9}{35}}},{\red {\frac{24}{35}}},\frac{61}{140}$  \\
$12_{\frac{173}{35}}$ & $67.26$ & ${\red {0}},\frac{1}{4},\frac{13}{20},{\red {\frac{2}{5}}},\frac{3}{28},{\red {\frac{6}{7}}},{\red {\frac{2}{7}}},\frac{15}{28},{\red {\frac{9}{35}}},\frac{71}{140},\frac{131}{140},{\red {\frac{24}{35}}}$  \\
$18_{\frac{68}{35}}$ & $100.8$ & ${\red {0}},\frac{2}{3},\frac{2}{3},\frac{1}{15},\frac{1}{15},{\red {\frac{2}{5}}},{\red {\frac{6}{7}}},\frac{11}{21},\frac{11}{21},\frac{20}{21},\frac{20}{21},{\red {\frac{2}{7}}},\frac{97}{105},\frac{97}{105},{\red {\frac{9}{35}}},{\red {\frac{24}{35}}},\frac{37}{105},\frac{37}{105}$  \\
$18_{\frac{208}{35}}$ & $100.8$ & ${\red {0}},\frac{1}{3},\frac{1}{3},\frac{11}{15},\frac{11}{15},{\red {\frac{2}{5}}},{\red {\frac{6}{7}}},\frac{4}{21},\frac{4}{21},{\red {\frac{2}{7}}},\frac{13}{21},\frac{13}{21},{\red {\frac{9}{35}}},\frac{62}{105},\frac{62}{105},\frac{2}{105},\frac{2}{105},{\red {\frac{24}{35}}}$  \\
\hline
\end{tabular}
}

~

\centerline{
\def\arraystretch{1.25} \setlength\tabcolsep{3pt}
\begin{tabular}{ |c|c|l|l| }
\hline
$N_{c}$ & $D^2$ & $s_1,s_2,\cdots$ \hfill \blue{$d_1,d_2,\cdots$} \\ 
\hline
$6_{\frac{142}{35}}$ & $33.63$ & ${\red {0}},{\red {\frac{3}{5}}},{\red {\frac{1}{7}}},{\red {\frac{5}{7}}},{\red {\frac{26}{35}}},{\red {\frac{11}{35}}}$ \hfill \blue{$1,\zeta_{3}^{1},\zeta_{5}^{1},\zeta_{5}^{2},2.915,3.635$} \\
$12_{\frac{177}{35}}$ & $67.26$ & ${\red {0}},\frac{1}{4},\frac{17}{20},{\red {\frac{3}{5}}},{\red {\frac{1}{7}}},\frac{11}{28},\frac{27}{28},{\red {\frac{5}{7}}},\frac{139}{140},{\red {\frac{26}{35}}},{\red {\frac{11}{35}}},\frac{79}{140}$  \\
$12_{\frac{107}{35}}$ & $67.26$ & ${\red {0}},\frac{3}{4},\frac{7}{20},{\red {\frac{3}{5}}},\frac{25}{28},{\red {\frac{1}{7}}},{\red {\frac{5}{7}}},\frac{13}{28},{\red {\frac{26}{35}}},\frac{69}{140},\frac{9}{140},{\red {\frac{11}{35}}}$  \\
$18_{\frac{212}{35}}$ & $100.8$ & ${\red {0}},\frac{1}{3},\frac{1}{3},\frac{14}{15},\frac{14}{15},{\red {\frac{3}{5}}},{\red {\frac{1}{7}}},\frac{10}{21},\frac{10}{21},\frac{1}{21},\frac{1}{21},{\red {\frac{5}{7}}},\frac{8}{105},\frac{8}{105},{\red {\frac{26}{35}}},{\red {\frac{11}{35}}},\frac{68}{105},\frac{68}{105}$  \\
$18_{\frac{72}{35}}$ & $100.8$ & ${\red {0}},\frac{2}{3},\frac{2}{3},\frac{4}{15},\frac{4}{15},{\red {\frac{3}{5}}},{\red {\frac{1}{7}}},\frac{17}{21},\frac{17}{21},{\red {\frac{5}{7}}},\frac{8}{21},\frac{8}{21},{\red {\frac{26}{35}}},\frac{43}{105},\frac{43}{105},\frac{103}{105},\frac{103}{105},{\red {\frac{11}{35}}}$  \\
\hline
\end{tabular}
}

~

\centerline{
\def\arraystretch{1.25} \setlength\tabcolsep{3pt}
\begin{tabular}{ |c|c|l|l| }
\hline
$N_{c}$ & $D^2$ & $s_1,s_2,\cdots$ \hfill \blue{$d_1,d_2,\cdots$} \\ 
\hline
$6_{\frac{222}{35}}$ & $33.63$ & ${\red {0}},{\red {\frac{3}{5}}},{\red {\frac{6}{7}}},{\red {\frac{2}{7}}},{\red {\frac{16}{35}}},{\red {\frac{31}{35}}}$ \hfill \blue{$1,\zeta_{3}^{1},\zeta_{5}^{1},\zeta_{5}^{2},2.915,3.635$} \\
$12_{\frac{257}{35}}$ & $67.26$ & ${\red {0}},\frac{1}{4},\frac{17}{20},{\red {\frac{3}{5}}},\frac{3}{28},{\red {\frac{6}{7}}},{\red {\frac{2}{7}}},\frac{15}{28},\frac{99}{140},{\red {\frac{16}{35}}},{\red {\frac{31}{35}}},\frac{19}{140}$  \\
$12_{\frac{187}{35}}$ & $67.26$ & ${\red {0}},\frac{3}{4},\frac{7}{20},{\red {\frac{3}{5}}},{\red {\frac{6}{7}}},\frac{17}{28},\frac{1}{28},{\red {\frac{2}{7}}},\frac{29}{140},{\red {\frac{16}{35}}},{\red {\frac{31}{35}}},\frac{89}{140}$  \\
$18_{\frac{12}{35}}$ & $100.8$ & ${\red {0}},\frac{1}{3},\frac{1}{3},\frac{14}{15},\frac{14}{15},{\red {\frac{3}{5}}},{\red {\frac{6}{7}}},\frac{4}{21},\frac{4}{21},{\red {\frac{2}{7}}},\frac{13}{21},\frac{13}{21},\frac{83}{105},\frac{83}{105},{\red {\frac{16}{35}}},{\red {\frac{31}{35}}},\frac{23}{105},\frac{23}{105}$  \\
$18_{\frac{152}{35}}$ & $100.8$ & ${\red {0}},\frac{2}{3},\frac{2}{3},\frac{4}{15},\frac{4}{15},{\red {\frac{3}{5}}},{\red {\frac{6}{7}}},\frac{11}{21},\frac{11}{21},\frac{20}{21},\frac{20}{21},{\red {\frac{2}{7}}},\frac{13}{105},\frac{13}{105},{\red {\frac{16}{35}}},{\red {\frac{31}{35}}},\frac{58}{105},\frac{58}{105}$  \\
\hline
\end{tabular}
}

\vfill\break

The following two non-Abelian families are obtained by stacking a FQH state
from the two Fibonacci non-Abelian families with a FQH state from the Ising
non-Abelian family.\\
\centerline{
\def\arraystretch{1.25} \setlength\tabcolsep{3pt}
\begin{tabular}{ |c|c|l|l| }
\hline
$N_{c}$ & $D^2$ & $s_1,s_2,\cdots$ \hfill \blue{$d_1,d_2,\cdots$} \\ 
\hline
$6_{\frac{3}{10}}$ & $14.47$ & ${\red {0}},{\red {\frac{1}{2}}},\frac{11}{16},{\red {\frac{9}{10}}},{\red {\frac{2}{5}}},\frac{7}{80}$ \hfill \blue{$1,1,\zeta_{2}^{1},\zeta_{3}^{1},\zeta_{3}^{1},2.288$} \\
$6_{\frac{73}{10}}$ & $14.47$ & ${\red {0}},{\red {\frac{1}{2}}},\frac{9}{16},{\red {\frac{9}{10}}},{\red {\frac{2}{5}}},\frac{77}{80}$  \\
$6_{\frac{13}{10}}$ & $14.47$ & ${\red {0}},{\red {\frac{1}{2}}},\frac{13}{16},{\red {\frac{9}{10}}},{\red {\frac{2}{5}}},\frac{17}{80}$  \\
$6_{\frac{63}{10}}$ & $14.47$ & ${\red {0}},{\red {\frac{1}{2}}},\frac{7}{16},{\red {\frac{9}{10}}},{\red {\frac{2}{5}}},\frac{67}{80}$  \\
$6_{\frac{23}{10}}$ & $14.47$ & ${\red {0}},{\red {\frac{1}{2}}},\frac{15}{16},{\red {\frac{9}{10}}},{\red {\frac{2}{5}}},\frac{27}{80}$  \\
$6_{\frac{53}{10}}$ & $14.47$ & ${\red {0}},{\red {\frac{1}{2}}},\frac{5}{16},{\red {\frac{9}{10}}},{\red {\frac{2}{5}}},\frac{57}{80}$  \\
$6_{\frac{33}{10}}$ & $14.47$ & ${\red {0}},{\red {\frac{1}{2}}},\frac{1}{16},{\red {\frac{9}{10}}},{\red {\frac{2}{5}}},\frac{37}{80}$  \\
\hline
\end{tabular}
}

~

\centerline{
\def\arraystretch{1.25} \setlength\tabcolsep{3pt}
\begin{tabular}{ |c|c|l|l| }
\hline
$N_{c}$ & $D^2$ & $s_1,s_2,\cdots$ \hfill \blue{$d_1,d_2,\cdots$} \\ 
\hline
$6_{\frac{77}{10}}$ & $14.47$ & ${\red {0}},{\red {\frac{1}{2}}},\frac{5}{16},{\red {\frac{1}{10}}},{\red {\frac{3}{5}}},\frac{73}{80}$ \hfill \blue{$1,1,\zeta_{2}^{1},\zeta_{3}^{1},\zeta_{3}^{1},2.288$} \\
$6_{\frac{7}{10}}$ & $14.47$ & ${\red {0}},{\red {\frac{1}{2}}},\frac{7}{16},{\red {\frac{1}{10}}},{\red {\frac{3}{5}}},\frac{3}{80}$  \\
$6_{\frac{67}{10}}$ & $14.47$ & ${\red {0}},{\red {\frac{1}{2}}},\frac{3}{16},{\red {\frac{1}{10}}},{\red {\frac{3}{5}}},\frac{63}{80}$  \\
$6_{\frac{17}{10}}$ & $14.47$ & ${\red {0}},{\red {\frac{1}{2}}},\frac{9}{16},{\red {\frac{1}{10}}},{\red {\frac{3}{5}}},\frac{13}{80}$  \\
$6_{\frac{57}{10}}$ & $14.47$ & ${\red {0}},{\red {\frac{1}{2}}},\frac{1}{16},{\red {\frac{1}{10}}},{\red {\frac{3}{5}}},\frac{53}{80}$  \\
$6_{\frac{27}{10}}$ & $14.47$ & ${\red {0}},{\red {\frac{1}{2}}},\frac{11}{16},{\red {\frac{1}{10}}},{\red {\frac{3}{5}}},\frac{23}{80}$  \\
$6_{\frac{47}{10}}$ & $14.47$ & ${\red {0}},{\red {\frac{1}{2}}},\frac{15}{16},{\red {\frac{1}{10}}},{\red {\frac{3}{5}}},\frac{43}{80}$  \\
\hline
\end{tabular}
}

\vfill
\break

$\mathfrak{so}(10)_2$ non-Abelian family:\\
\centerline{
\def\arraystretch{1.25} \setlength\tabcolsep{3pt}
\begin{tabular}{ |c|c|l|l| }
\hline
$N_{c}$ & $D^2$ & $s_1,s_2,\cdots$ \hfill \blue{$d_1,d_2,\cdots$} \\ 
\hline
$6_{0}$ & $20$ & ${\red {0}},{\red {0}},{\red {\frac{1}{5}}},{\red {\frac{4}{5}}},0,\frac{1}{2}$ \hfill \blue{$1,1,2,2,\sqrt{5},\sqrt{5}$} \\
$6_{0}$ & $20$ & ${\red {0}},{\red {0}},{\red {\frac{1}{5}}},{\red {\frac{4}{5}}},\frac{1}{4},\frac{3}{4}$  \\
$12_{1}$ & $40$ & ${\red {0}},{\red {0}},\frac{1}{4},\frac{1}{4},\frac{1}{20},{\red {\frac{1}{5}}},{\red {\frac{4}{5}}},\frac{9}{20},0,\frac{1}{4},\frac{3}{4},\frac{1}{2}$  \\
$12_{1}$ & $40$ & ${\red {0}},{\red {0}},\frac{1}{4},\frac{1}{4},\frac{1}{20},{\red {\frac{1}{5}}},{\red {\frac{4}{5}}},\frac{9}{20},\frac{1}{16},\frac{1}{16},\frac{9}{16},\frac{9}{16}$  \\
$12_{1}$ & $40$ & ${\red {0}},{\red {0}},\frac{1}{4},\frac{1}{4},\frac{1}{20},{\red {\frac{1}{5}}},{\red {\frac{4}{5}}},\frac{9}{20},\frac{13}{16},\frac{13}{16},\frac{5}{16},\frac{5}{16}$  \\
$12_{7}$ & $40$ & ${\red {0}},{\red {0}},\frac{3}{4},\frac{3}{4},\frac{19}{20},{\red {\frac{1}{5}}},{\red {\frac{4}{5}}},\frac{11}{20},0,\frac{1}{4},\frac{3}{4},\frac{1}{2}$  \\
$12_{7}$ & $40$ & ${\red {0}},{\red {0}},\frac{3}{4},\frac{3}{4},\frac{19}{20},{\red {\frac{1}{5}}},{\red {\frac{4}{5}}},\frac{11}{20},\frac{15}{16},\frac{15}{16},\frac{7}{16},\frac{7}{16}$  \\
\hline
\end{tabular}
}

~

$\mathfrak{so}(5)_2$ non-Abelian family:\\
\centerline{
\def\arraystretch{1.25} \setlength\tabcolsep{3pt}
\begin{tabular}{ |c|c|l|l| }
\hline
$N_{c}$ & $D^2$ & $s_1,s_2,\cdots$ \hfill \blue{$d_1,d_2,\cdots$} \\ 
\hline
$6_{4}$ & $20$ & ${\red {0}},{\red {0}},{\red {\frac{2}{5}}},{\red {\frac{3}{5}}},\frac{1}{4},\frac{3}{4}$ \hfill \blue{$1,1,2,2,\sqrt{5},\sqrt{5}$} \\
$6_{4}$ & $20$ & ${\red {0}},{\red {0}},{\red {\frac{2}{5}}},{\red {\frac{3}{5}}},0,\frac{1}{2}$  \\
$12_{3}$ & $40$ & ${\red {0}},{\red {0}},\frac{3}{4},\frac{3}{4},\frac{3}{20},\frac{7}{20},{\red {\frac{2}{5}}},{\red {\frac{3}{5}}},0,\frac{1}{4},\frac{3}{4},\frac{1}{2}$  \\
$12_{3}$ & $40$ & ${\red {0}},{\red {0}},\frac{3}{4},\frac{3}{4},\frac{3}{20},\frac{7}{20},{\red {\frac{2}{5}}},{\red {\frac{3}{5}}},\frac{3}{16},\frac{3}{16},\frac{11}{16},\frac{11}{16}$  \\
$12_{3}$ & $40$ & ${\red {0}},{\red {0}},\frac{3}{4},\frac{3}{4},\frac{3}{20},\frac{7}{20},{\red {\frac{2}{5}}},{\red {\frac{3}{5}}},\frac{15}{16},\frac{15}{16},\frac{7}{16},\frac{7}{16}$  \\
$12_{5}$ & $40$ & ${\red {0}},{\red {0}},\frac{1}{4},\frac{1}{4},\frac{17}{20},\frac{13}{20},{\red {\frac{2}{5}}},{\red {\frac{3}{5}}},0,\frac{1}{4},\frac{3}{4},\frac{1}{2}$  \\
$12_{5}$ & $40$ & ${\red {0}},{\red {0}},\frac{1}{4},\frac{1}{4},\frac{17}{20},\frac{13}{20},{\red {\frac{2}{5}}},{\red {\frac{3}{5}}},\frac{13}{16},\frac{13}{16},\frac{5}{16},\frac{5}{16}$  \\
\hline
\end{tabular}
}

~

The following $SU(2)_{11}$ non-Abelian family contains FQH state \cite{Wnab,W9927}
\begin{align}
\Psi_{\nu=-\frac {11}2}(\{\bar z_i\})=[\chi_{11}(\{\bar z_i\})]^2,\ \ N_c=12_{\frac {59}{13}} 
\end{align}
\centerline{
\def\arraystretch{1.25} \setlength\tabcolsep{3pt}
\begin{tabular}{ |c|c|l|l| }
\hline
$N_{c}$ & $D^2$ & $s_1,s_2,\cdots$ \hfill \blue{$d_1,d_2,\cdots$} \\ 
\hline
$6_{\frac{46}{13}}$ & $56.74$ & ${\red {0}},{\red {\frac{4}{13}}},{\red {\frac{2}{13}}},{\red {\frac{7}{13}}},{\red {\frac{6}{13}}},{\red {\frac{12}{13}}}$ \hfill \blue{$1,\zeta_{11}^{1},\zeta_{11}^{2},\zeta_{11}^{3},\zeta_{11}^{4},\zeta_{11}^{5}$} \\
$12_{\frac{33}{13}}$ & $113.4$ & ${\red {0}},\frac{3}{4},\frac{3}{52},{\red {\frac{4}{13}}},\frac{47}{52},{\red {\frac{2}{13}}},\frac{15}{52},{\red {\frac{7}{13}}},\frac{11}{52},{\red {\frac{6}{13}}},{\red {\frac{12}{13}}},\frac{35}{52}$  \\
$\red 12_{\frac{59}{13}}$ & $113.4$ & ${\red {0}},\frac{1}{4},{\red {\frac{4}{13}}},\frac{29}{52},{\red {\frac{2}{13}}},\frac{21}{52},\frac{41}{52},{\red {\frac{7}{13}}},\frac{37}{52},{\red {\frac{6}{13}}},{\red {\frac{12}{13}}},\frac{9}{52}$  \\
$18_{\frac{20}{13}}$ & $170.2$ & ${\red {0}},\frac{2}{3},\frac{2}{3},\frac{38}{39},\frac{38}{39},{\red {\frac{4}{13}}},{\red {\frac{2}{13}}},\frac{32}{39},\frac{32}{39},\frac{8}{39},\frac{8}{39},{\red {\frac{7}{13}}},\frac{5}{39},\frac{5}{39},{\red {\frac{6}{13}}},{\red {\frac{12}{13}}},\frac{23}{39},\frac{23}{39}$  \\
$18_{\frac{72}{13}}$ & $170.2$ & ${\red {0}},\frac{1}{3},\frac{1}{3},{\red {\frac{4}{13}}},\frac{25}{39},\frac{25}{39},{\red {\frac{2}{13}}},\frac{19}{39},\frac{19}{39},\frac{34}{39},\frac{34}{39},{\red {\frac{7}{13}}},\frac{31}{39},\frac{31}{39},{\red {\frac{6}{13}}},{\red {\frac{12}{13}}},\frac{10}{39},\frac{10}{39}$  \\
\hline
\end{tabular}
}

~

The following $SU(2)_{-11}$ non-Abelian family contains FQH state \cite{Wnab,W9927}
\begin{align}
\Psi_{\nu=\frac {11}2}(\{z_i\})=[\chi_{11}(\{z_i\})]^2,\ \ N_c=12_{\frac {45}{13}} 
\end{align}
\centerline{
\def\arraystretch{1.25} \setlength\tabcolsep{3pt}
\begin{tabular}{ |c|c|l|l| }
\hline
$N_{c}$ & $D^2$ & $s_1,s_2,\cdots$ \hfill \blue{$d_1,d_2,\cdots$} \\ 
\hline
$6_{\frac{58}{13}}$ & $56.74$ & ${\red {0}},{\red {\frac{9}{13}}},{\red {\frac{11}{13}}},{\red {\frac{6}{13}}},{\red {\frac{7}{13}}},{\red {\frac{1}{13}}}$ \hfill \blue{$1,\zeta_{11}^{1},\zeta_{11}^{2},\zeta_{11}^{3},\zeta_{11}^{4},\zeta_{11}^{5}$} \\
$12_{\frac{71}{13}}$ & $113.4$ & ${\red {0}},\frac{1}{4},\frac{49}{52},{\red {\frac{9}{13}}},\frac{5}{52},{\red {\frac{11}{13}}},\frac{37}{52},{\red {\frac{6}{13}}},\frac{41}{52},{\red {\frac{7}{13}}},{\red {\frac{1}{13}}},\frac{17}{52}$  \\
$\red 12_{\frac{45}{13}}$ & $113.4$ & ${\red {0}},\frac{3}{4},{\red {\frac{9}{13}}},\frac{23}{52},{\red {\frac{11}{13}}},\frac{31}{52},\frac{11}{52},{\red {\frac{6}{13}}},\frac{15}{52},{\red {\frac{7}{13}}},{\red {\frac{1}{13}}},\frac{43}{52}$  \\
$18_{\frac{84}{13}}$ & $170.2$ & ${\red {0}},\frac{1}{3},\frac{1}{3},\frac{1}{39},\frac{1}{39},{\red {\frac{9}{13}}},{\red {\frac{11}{13}}},\frac{7}{39},\frac{7}{39},\frac{31}{39},\frac{31}{39},{\red {\frac{6}{13}}},\frac{34}{39},\frac{34}{39},{\red {\frac{7}{13}}},{\red {\frac{1}{13}}},\frac{16}{39},\frac{16}{39}$  \\
$18_{\frac{32}{13}}$ & $170.2$ & ${\red {0}},\frac{2}{3},\frac{2}{3},{\red {\frac{9}{13}}},\frac{14}{39},\frac{14}{39},{\red {\frac{11}{13}}},\frac{20}{39},\frac{20}{39},\frac{5}{39},\frac{5}{39},{\red {\frac{6}{13}}},\frac{8}{39},\frac{8}{39},{\red {\frac{7}{13}}},{\red {\frac{1}{13}}},\frac{29}{39},\frac{29}{39}$  \\
\hline
\end{tabular}
}

~

The following is the $\mathfrak{so}(8)_{-3}$ non-Abelian family:

\centerline{
\def\arraystretch{1.25} \setlength\tabcolsep{3pt}
\begin{tabular}{ |c|c|l|l| }
\hline
$N_{c}$ & $D^2$ & $s_1,s_2,\cdots$ \hfill \blue{$d_1,d_2,\cdots$} \\ 
\hline
$6_{\frac{8}{3}}$ & $74.61$ & ${\red {0}},{\red {\frac{1}{9}}},{\red {\frac{1}{9}}},{\red {\frac{1}{9}}},{\red {\frac{1}{3}}},{\red {\frac{2}{3}}}$ \hfill \blue{$1,\zeta_{7}^{3},\zeta_{7}^{3},\zeta_{7}^{3},\zeta_{16}^{4},\zeta_{16}^{6}$} \\
$12_{\frac{5}{3}}$ & $149.2$ & ${\red {0}},\frac{3}{4},{\red {\frac{1}{9}}},{\red {\frac{1}{9}}},{\red {\frac{1}{9}}},\frac{31}{36},\frac{31}{36},\frac{31}{36},\frac{1}{12},{\red {\frac{1}{3}}},{\red {\frac{2}{3}}},\frac{5}{12}$  \\
$12_{\frac{11}{3}}$ & $149.2$ & ${\red {0}},\frac{1}{4},{\red {\frac{1}{9}}},{\red {\frac{1}{9}}},{\red {\frac{1}{9}}},\frac{13}{36},\frac{13}{36},\frac{13}{36},{\red {\frac{1}{3}}},\frac{7}{12},\frac{11}{12},{\red {\frac{2}{3}}}$  \\
$18_{\frac{2}{3}}$ & $223.8$ & ${\red {0}},\frac{2}{3},\frac{2}{3},{\red {\frac{1}{9}}},{\red {\frac{1}{9}}},{\red {\frac{1}{9}}},\frac{7}{9},\frac{7}{9},\frac{7}{9},\frac{7}{9},\frac{7}{9},\frac{7}{9},0,0,{\red {\frac{1}{3}}},\frac{1}{3},\frac{1}{3},{\red {\frac{2}{3}}}$  \\
$18_{\frac{14}{3}}$ & $223.8$ & ${\red {0}},\frac{1}{3},\frac{1}{3},{\red {\frac{1}{9}}},{\red {\frac{1}{9}}},{\red {\frac{1}{9}}},\frac{4}{9},\frac{4}{9},\frac{4}{9},\frac{4}{9},\frac{4}{9},\frac{4}{9},{\red {\frac{1}{3}}},\frac{2}{3},\frac{2}{3},0,0,{\red {\frac{2}{3}}}$  \\
\hline
\end{tabular}
}

~

The following is the $\mathfrak{so}(8)_{3}$ non-Abelian family:

\centerline{
\def\arraystretch{1.25} \setlength\tabcolsep{3pt}
\begin{tabular}{ |c|c|l|l| }
\hline
$N_{c}$ & $D^2$ & $s_1,s_2,\cdots$ \hfill \blue{$d_1,d_2,\cdots$} \\ 
\hline
$6_{\frac{16}{3}}$ & $74.61$ & ${\red {0}},{\red {\frac{8}{9}}},{\red {\frac{8}{9}}},{\red {\frac{8}{9}}},{\red {\frac{2}{3}}},{\red {\frac{1}{3}}}$ \hfill \blue{$1,\zeta_{7}^{3},\zeta_{7}^{3},\zeta_{7}^{3},\zeta_{16}^{4},\zeta_{16}^{6}$} \\
$12_{\frac{19}{3}}$ & $149.2$ & ${\red {0}},\frac{1}{4},{\red {\frac{8}{9}}},{\red {\frac{8}{9}}},{\red {\frac{8}{9}}},\frac{5}{36},\frac{5}{36},\frac{5}{36},\frac{11}{12},{\red {\frac{2}{3}}},{\red {\frac{1}{3}}},\frac{7}{12}$  \\
$12_{\frac{13}{3}}$ & $149.2$ & ${\red {0}},\frac{3}{4},{\red {\frac{8}{9}}},{\red {\frac{8}{9}}},{\red {\frac{8}{9}}},\frac{23}{36},\frac{23}{36},\frac{23}{36},{\red {\frac{2}{3}}},\frac{5}{12},\frac{1}{12},{\red {\frac{1}{3}}}$  \\
$18_{\frac{22}{3}}$ & $223.8$ & ${\red {0}},\frac{1}{3},\frac{1}{3},{\red {\frac{8}{9}}},{\red {\frac{8}{9}}},{\red {\frac{8}{9}}},\frac{2}{9},\frac{2}{9},\frac{2}{9},\frac{2}{9},\frac{2}{9},\frac{2}{9},0,0,{\red {\frac{2}{3}}},{\red {\frac{1}{3}}},\frac{2}{3},\frac{2}{3}$  \\
$18_{\frac{10}{3}}$ & $223.8$ & ${\red {0}},\frac{2}{3},\frac{2}{3},{\red {\frac{8}{9}}},{\red {\frac{8}{9}}},{\red {\frac{8}{9}}},\frac{5}{9},\frac{5}{9},\frac{5}{9},\frac{5}{9},\frac{5}{9},\frac{5}{9},\frac{1}{3},\frac{1}{3},{\red {\frac{2}{3}}},0,0,{\red {\frac{1}{3}}}$  \\
\hline
\end{tabular}
}

~

The following is the $(E_7)_{3}$ non-Abelian family:

\centerline{
\def\arraystretch{1.25} \setlength\tabcolsep{3pt}
\begin{tabular}{ |c|c|l|l| }
\hline
$N_{c}$ & $D^2$ & $s_1,s_2,\cdots$ \hfill \blue{$d_1,d_2,\cdots$} \\ 
\hline
$6_{2}$ & $100.6$ & ${\red {0}},{\red {\frac{6}{7}}},{\red {\frac{5}{7}}},{\red {\frac{3}{7}}},{\red {0}},{\red {\frac{1}{3}}}$ \hfill \blue{$1,\frac{3+\sqrt{21}}{2},\frac{3+\sqrt{21}}{2},\frac{3+\sqrt{21}}{2},\frac{5+\sqrt{21}}{2},\frac{7+\sqrt{21}}{2}$} \\
$12_{1}$ & $201.2$ & ${\red {0}},\frac{3}{4},{\red {\frac{6}{7}}},\frac{5}{28},{\red {\frac{5}{7}}},\frac{17}{28},{\red {\frac{3}{7}}},\frac{13}{28},{\red {0}},\frac{3}{4},\frac{1}{12},{\red {\frac{1}{3}}}$  \\
$12_{3}$ & $201.2$ & ${\red {0}},\frac{1}{4},\frac{27}{28},\frac{3}{28},{\red {\frac{6}{7}}},{\red {\frac{5}{7}}},\frac{19}{28},{\red {\frac{3}{7}}},{\red {0}},\frac{1}{4},{\red {\frac{1}{3}}},\frac{7}{12}$  \\
$18_{0}$ & $301.8$ & ${\red {0}},\frac{2}{3},\frac{2}{3},\frac{2}{21},\frac{2}{21},{\red {\frac{6}{7}}},{\red {\frac{5}{7}}},\frac{8}{21},\frac{8}{21},{\red {\frac{3}{7}}},\frac{11}{21},\frac{11}{21},{\red {0}},\frac{2}{3},\frac{2}{3},0,0,{\red {\frac{1}{3}}}$  \\
$18_{4}$ & $301.8$ & ${\red {0}},\frac{1}{3},\frac{1}{3},\frac{1}{21},\frac{1}{21},{\red {\frac{6}{7}}},\frac{4}{21},\frac{4}{21},\frac{16}{21},\frac{16}{21},{\red {\frac{5}{7}}},{\red {\frac{3}{7}}},{\red {0}},\frac{1}{3},\frac{1}{3},{\red {\frac{1}{3}}},\frac{2}{3},\frac{2}{3}$  \\
\hline
\end{tabular}
}

~

The following is the $(E_7)_{-3}$ non-Abelian family:

\centerline{
\def\arraystretch{1.25} \setlength\tabcolsep{3pt}
\begin{tabular}{ |c|c|l|l| }
\hline
$N_{c}$ & $D^2$ & $s_1,s_2,\cdots$ \hfill \blue{$d_1,d_2,\cdots$} \\ 
\hline
$6_{6}$ & $100.6$ & ${\red {0}},{\red {\frac{1}{7}}},{\red {\frac{2}{7}}},{\red {\frac{4}{7}}},{\red {0}},{\red {\frac{2}{3}}}$ \hfill \blue{$1,\frac{3+\sqrt{21}}{2},\frac{3+\sqrt{21}}{2},\frac{3+\sqrt{21}}{2},\frac{5+\sqrt{21}}{2},\frac{7+\sqrt{21}}{2}$} \\
$12_{7}$ & $201.2$ & ${\red {0}},\frac{1}{4},{\red {\frac{1}{7}}},\frac{23}{28},{\red {\frac{2}{7}}},\frac{11}{28},{\red {\frac{4}{7}}},\frac{15}{28},{\red {0}},\frac{1}{4},\frac{11}{12},{\red {\frac{2}{3}}}$  \\
$12_{5}$ & $201.2$ & ${\red {0}},\frac{3}{4},\frac{1}{28},\frac{25}{28},{\red {\frac{1}{7}}},{\red {\frac{2}{7}}},\frac{9}{28},{\red {\frac{4}{7}}},{\red {0}},\frac{3}{4},{\red {\frac{2}{3}}},\frac{5}{12}$  \\
$18_{0}$ & $301.8$ & ${\red {0}},\frac{1}{3},\frac{1}{3},\frac{19}{21},\frac{19}{21},{\red {\frac{1}{7}}},{\red {\frac{2}{7}}},\frac{13}{21},\frac{13}{21},{\red {\frac{4}{7}}},\frac{10}{21},\frac{10}{21},{\red {0}},\frac{1}{3},\frac{1}{3},0,0,{\red {\frac{2}{3}}}$  \\
$18_{4}$ & $301.8$ & ${\red {0}},\frac{2}{3},\frac{2}{3},\frac{20}{21},\frac{20}{21},{\red {\frac{1}{7}}},\frac{17}{21},\frac{17}{21},\frac{5}{21},\frac{5}{21},{\red {\frac{2}{7}}},{\red {\frac{4}{7}}},{\red {0}},\frac{2}{3},\frac{2}{3},\frac{1}{3},\frac{1}{3},{\red {\frac{2}{3}}}$  \\
\hline
\end{tabular}
}

\medskip The following are other non-Abelian families:\medskip

\centerline{
\def\arraystretch{1.25} \setlength\tabcolsep{3pt}
\begin{tabular}{ |c|c|l|l| }
\hline
$N_{c}$ & $D^2$ & $s_1,s_2,\cdots$ \hfill \blue{$d_1,d_2,\cdots$} \\ 
\hline
$7_{\frac{1}{4}}$ & $27.31$ & ${\red {0}},{\red {\frac{1}{2}}},\frac{27}{32},\frac{27}{32},{\red {\frac{1}{4}}},{\red {\frac{3}{4}}},\frac{7}{32}$ \hfill \blue{$1,1,\zeta_{6}^{1},\zeta_{6}^{1},\zeta_{6}^{2},\zeta_{6}^{2},\zeta_{6}^{3}$} \\
$7_{\frac{29}{4}}$ & $27.31$ & ${\red {0}},{\red {\frac{1}{2}}},\frac{23}{32},\frac{23}{32},{\red {\frac{1}{4}}},{\red {\frac{3}{4}}},\frac{3}{32}$  \\
$7_{\frac{5}{4}}$ & $27.31$ & ${\red {0}},{\red {\frac{1}{2}}},\frac{31}{32},\frac{31}{32},{\red {\frac{1}{4}}},{\red {\frac{3}{4}}},\frac{11}{32}$  \\
$7_{\frac{25}{4}}$ & $27.31$ & ${\red {0}},{\red {\frac{1}{2}}},\frac{19}{32},\frac{19}{32},{\red {\frac{1}{4}}},{\red {\frac{3}{4}}},\frac{31}{32}$  \\
$7_{\frac{9}{4}}$ & $27.31$ & ${\red {0}},{\red {\frac{1}{2}}},\frac{3}{32},\frac{3}{32},{\red {\frac{1}{4}}},{\red {\frac{3}{4}}},\frac{15}{32}$  \\
$7_{\frac{21}{4}}$ & $27.31$ & ${\red {0}},{\red {\frac{1}{2}}},\frac{15}{32},\frac{15}{32},{\red {\frac{1}{4}}},{\red {\frac{3}{4}}},\frac{27}{32}$  \\
$7_{\frac{13}{4}}$ & $27.31$ & ${\red {0}},{\red {\frac{1}{2}}},\frac{7}{32},\frac{7}{32},{\red {\frac{1}{4}}},{\red {\frac{3}{4}}},\frac{19}{32}$  \\
\hline
\end{tabular}
}

~

\centerline{
\def\arraystretch{1.25} \setlength\tabcolsep{3pt}
\begin{tabular}{ |c|c|l|l| }
\hline
$N_{c}$ & $D^2$ & $s_1,s_2,\cdots$ \hfill \blue{$d_1,d_2,\cdots$} \\ 
\hline
$7_{\frac{31}{4}}$ & $27.31$ & ${\red {0}},{\red {\frac{1}{2}}},\frac{5}{32},\frac{5}{32},{\red {\frac{1}{4}}},{\red {\frac{3}{4}}},\frac{25}{32}$ \hfill \blue{$1,1,\zeta_{6}^{1},\zeta_{6}^{1},\zeta_{6}^{2},\zeta_{6}^{2},\zeta_{6}^{3}$} \\
$7_{\frac{3}{4}}$ & $27.31$ & ${\red {0}},{\red {\frac{1}{2}}},\frac{9}{32},\frac{9}{32},{\red {\frac{1}{4}}},{\red {\frac{3}{4}}},\frac{29}{32}$  \\
$7_{\frac{27}{4}}$ & $27.31$ & ${\red {0}},{\red {\frac{1}{2}}},\frac{1}{32},\frac{1}{32},{\red {\frac{1}{4}}},{\red {\frac{3}{4}}},\frac{21}{32}$  \\
$7_{\frac{7}{4}}$ & $27.31$ & ${\red {0}},{\red {\frac{1}{2}}},\frac{13}{32},\frac{13}{32},{\red {\frac{1}{4}}},{\red {\frac{3}{4}}},\frac{1}{32}$  \\
$7_{\frac{23}{4}}$ & $27.31$ & ${\red {0}},{\red {\frac{1}{2}}},\frac{29}{32},\frac{29}{32},{\red {\frac{1}{4}}},{\red {\frac{3}{4}}},\frac{17}{32}$  \\
$7_{\frac{11}{4}}$ & $27.31$ & ${\red {0}},{\red {\frac{1}{2}}},\frac{17}{32},\frac{17}{32},{\red {\frac{1}{4}}},{\red {\frac{3}{4}}},\frac{5}{32}$  \\
$7_{\frac{19}{4}}$ & $27.31$ & ${\red {0}},{\red {\frac{1}{2}}},\frac{25}{32},\frac{25}{32},{\red {\frac{1}{4}}},{\red {\frac{3}{4}}},\frac{13}{32}$  \\
\hline
\end{tabular}
}

~

\centerline{
\def\arraystretch{1.25} \setlength\tabcolsep{3pt}
\begin{tabular}{ |c|c|l|l| }
\hline
$N_{c}$ & $D^2$ & $s_1,s_2,\cdots$ \hfill \blue{$d_1,d_2,\cdots$} \\ 
\hline
$7_{2}$ & $28$ & ${\red {0}},{\red {0}},{\red {\frac{1}{7}}},{\red {\frac{2}{7}}},{\red {\frac{4}{7}}},\frac{1}{8},\frac{5}{8}$ \hfill \blue{$1,1,2,2,2,\sqrt{7},\sqrt{7}$} \\
$7_{2}$ & $28$ & ${\red {0}},{\red {0}},{\red {\frac{1}{7}}},{\red {\frac{2}{7}}},{\red {\frac{4}{7}}},\frac{7}{8},\frac{3}{8}$  \\
$14_{1}$ & $56$ & ${\red {0}},{\red {0}},\frac{3}{4},\frac{3}{4},\frac{1}{28},\frac{25}{28},{\red {\frac{1}{7}}},{\red {\frac{2}{7}}},\frac{9}{28},{\red {\frac{4}{7}}},\frac{1}{8},\frac{7}{8},\frac{3}{8},\frac{5}{8}$  \\
$14_{1}$ & $56$ & ${\red {0}},{\red {0}},\frac{3}{4},\frac{3}{4},\frac{1}{28},\frac{25}{28},{\red {\frac{1}{7}}},{\red {\frac{2}{7}}},\frac{9}{28},{\red {\frac{4}{7}}},\frac{1}{16},\frac{1}{16},\frac{9}{16},\frac{9}{16}$  \\
$14_{1}$ & $56$ & ${\red {0}},{\red {0}},\frac{3}{4},\frac{3}{4},\frac{1}{28},\frac{25}{28},{\red {\frac{1}{7}}},{\red {\frac{2}{7}}},\frac{9}{28},{\red {\frac{4}{7}}},\frac{13}{16},\frac{13}{16},\frac{5}{16},\frac{5}{16}$  \\
$14_{3}$ & $56$ & ${\red {0}},{\red {0}},\frac{1}{4},\frac{1}{4},{\red {\frac{1}{7}}},\frac{23}{28},{\red {\frac{2}{7}}},\frac{11}{28},{\red {\frac{4}{7}}},\frac{15}{28},\frac{1}{8},\frac{7}{8},\frac{3}{8},\frac{5}{8}$  \\
$14_{3}$ & $56$ & ${\red {0}},{\red {0}},\frac{1}{4},\frac{1}{4},{\red {\frac{1}{7}}},\frac{23}{28},{\red {\frac{2}{7}}},\frac{11}{28},{\red {\frac{4}{7}}},\frac{15}{28},\frac{3}{16},\frac{3}{16},\frac{11}{16},\frac{11}{16}$  \\
\hline
\end{tabular}
}

~

\centerline{
\def\arraystretch{1.25} \setlength\tabcolsep{3pt}
\begin{tabular}{ |c|c|l|l| }
\hline
$N_{c}$ & $D^2$ & $s_1,s_2,\cdots$ \hfill \blue{$d_1,d_2,\cdots$} \\ 
\hline
$7_{6}$ & $28$ & ${\red {0}},{\red {0}},{\red {\frac{6}{7}}},{\red {\frac{5}{7}}},{\red {\frac{3}{7}}},\frac{7}{8},\frac{3}{8}$ \hfill \blue{$1,1,2,2,2,\sqrt{7},\sqrt{7}$} \\
$7_{6}$ & $28$ & ${\red {0}},{\red {0}},{\red {\frac{6}{7}}},{\red {\frac{5}{7}}},{\red {\frac{3}{7}}},\frac{1}{8},\frac{5}{8}$  \\
$14_{7}$ & $56$ & ${\red {0}},{\red {0}},\frac{1}{4},\frac{1}{4},\frac{27}{28},\frac{3}{28},{\red {\frac{6}{7}}},{\red {\frac{5}{7}}},\frac{19}{28},{\red {\frac{3}{7}}},\frac{1}{8},\frac{7}{8},\frac{3}{8},\frac{5}{8}$  \\
$14_{7}$ & $56$ & ${\red {0}},{\red {0}},\frac{1}{4},\frac{1}{4},\frac{27}{28},\frac{3}{28},{\red {\frac{6}{7}}},{\red {\frac{5}{7}}},\frac{19}{28},{\red {\frac{3}{7}}},\frac{15}{16},\frac{15}{16},\frac{7}{16},\frac{7}{16}$  \\
$14_{7}$ & $56$ & ${\red {0}},{\red {0}},\frac{1}{4},\frac{1}{4},\frac{27}{28},\frac{3}{28},{\red {\frac{6}{7}}},{\red {\frac{5}{7}}},\frac{19}{28},{\red {\frac{3}{7}}},\frac{3}{16},\frac{3}{16},\frac{11}{16},\frac{11}{16}$  \\
$14_{5}$ & $56$ & ${\red {0}},{\red {0}},\frac{3}{4},\frac{3}{4},{\red {\frac{6}{7}}},\frac{5}{28},{\red {\frac{5}{7}}},\frac{17}{28},{\red {\frac{3}{7}}},\frac{13}{28},\frac{1}{8},\frac{7}{8},\frac{3}{8},\frac{5}{8}$  \\
$14_{5}$ & $56$ & ${\red {0}},{\red {0}},\frac{3}{4},\frac{3}{4},{\red {\frac{6}{7}}},\frac{5}{28},{\red {\frac{5}{7}}},\frac{17}{28},{\red {\frac{3}{7}}},\frac{13}{28},\frac{13}{16},\frac{13}{16},\frac{5}{16},\frac{5}{16}$  \\
\hline
\end{tabular}
}

~

\centerline{
\def\arraystretch{1.25} \setlength\tabcolsep{3pt}
\begin{tabular}{ |c|c|l|l| }
\hline
$N_{c}$ & $D^2$ & $s_1,s_2,\cdots$ \hfill \blue{$d_1,d_2,\cdots$} \\ 
\hline
$7_{\frac{8}{5}}$ & $86.75$ & ${\red {0}},{\red {\frac{4}{5}}},{\red {\frac{2}{15}}},{\red {0}},{\red {\frac{2}{5}}},{\red {\frac{1}{3}}},{\red {\frac{4}{5}}}$ \hfill \blue{$1,\zeta_{13}^{1},\zeta_{13}^{2},\zeta_{13}^{3},\zeta_{13}^{4},\zeta_{13}^{5},\zeta_{13}^{6}$} \\
$14_{\frac{3}{5}}$ & $173.5$ & ${\red {0}},\frac{3}{4},{\red {\frac{4}{5}}},\frac{11}{20},\frac{53}{60},{\red {\frac{2}{15}}},{\red {0}},\frac{3}{4},\frac{3}{20},{\red {\frac{2}{5}}},\frac{1}{12},{\red {\frac{1}{3}}},{\red {\frac{4}{5}}},\frac{11}{20}$  \\
$14_{\frac{13}{5}}$ & $173.5$ & ${\red {0}},\frac{1}{4},\frac{1}{20},{\red {\frac{4}{5}}},{\red {\frac{2}{15}}},\frac{23}{60},{\red {0}},\frac{1}{4},\frac{13}{20},{\red {\frac{2}{5}}},{\red {\frac{1}{3}}},\frac{7}{12},\frac{1}{20},{\red {\frac{4}{5}}}$  \\
$21_{\frac{38}{5}}$ & $260.2$ & ${\red {0}},\frac{2}{3},\frac{2}{3},{\red {\frac{4}{5}}},\frac{7}{15},\frac{7}{15},{\red {\frac{2}{15}}},\frac{4}{5},\frac{4}{5},{\red {0}},\frac{2}{3},\frac{2}{3},\frac{1}{15},\frac{1}{15},{\red {\frac{2}{5}}},0,0,{\red {\frac{1}{3}}},{\red {\frac{4}{5}}},\frac{7}{15},\frac{7}{15}$  \\
$21_{\frac{18}{5}}$ & $260.2$ & ${\red {0}},\frac{1}{3},\frac{1}{3},\frac{2}{15},\frac{2}{15},{\red {\frac{4}{5}}},{\red {\frac{2}{15}}},\frac{7}{15},\frac{7}{15},{\red {0}},\frac{1}{3},\frac{1}{3},\frac{11}{15},\frac{11}{15},{\red {\frac{2}{5}}},{\red {\frac{1}{3}}},\frac{2}{3},\frac{2}{3},\frac{2}{15},\frac{2}{15},{\red {\frac{4}{5}}}$  \\
\hline
\end{tabular}
}

~

\centerline{
\def\arraystretch{1.25} \setlength\tabcolsep{3pt}
\begin{tabular}{ |c|c|l|l| }
\hline
$N_{c}$ & $D^2$ & $s_1,s_2,\cdots$ \hfill \blue{$d_1,d_2,\cdots$} \\ 
\hline
$7_{\frac{32}{5}}$ & $86.75$ & ${\red {0}},{\red {\frac{1}{5}}},{\red {\frac{13}{15}}},{\red {0}},{\red {\frac{3}{5}}},{\red {\frac{2}{3}}},{\red {\frac{1}{5}}}$ \hfill \blue{$1,\zeta_{13}^{1},\zeta_{13}^{2},\zeta_{13}^{3},\zeta_{13}^{4},\zeta_{13}^{5},\zeta_{13}^{6}$} \\
$14_{\frac{37}{5}}$ & $173.5$ & ${\red {0}},\frac{1}{4},{\red {\frac{1}{5}}},\frac{9}{20},\frac{7}{60},{\red {\frac{13}{15}}},{\red {0}},\frac{1}{4},\frac{17}{20},{\red {\frac{3}{5}}},\frac{11}{12},{\red {\frac{2}{3}}},{\red {\frac{1}{5}}},\frac{9}{20}$  \\
$14_{\frac{27}{5}}$ & $173.5$ & ${\red {0}},\frac{3}{4},\frac{19}{20},{\red {\frac{1}{5}}},{\red {\frac{13}{15}}},\frac{37}{60},{\red {0}},\frac{3}{4},\frac{7}{20},{\red {\frac{3}{5}}},{\red {\frac{2}{3}}},\frac{5}{12},\frac{19}{20},{\red {\frac{1}{5}}}$  \\
$21_{\frac{2}{5}}$ & $260.2$ & ${\red {0}},\frac{1}{3},\frac{1}{3},{\red {\frac{1}{5}}},\frac{8}{15},\frac{8}{15},{\red {\frac{13}{15}}},\frac{1}{5},\frac{1}{5},{\red {0}},\frac{1}{3},\frac{1}{3},\frac{14}{15},\frac{14}{15},{\red {\frac{3}{5}}},0,0,{\red {\frac{2}{3}}},{\red {\frac{1}{5}}},\frac{8}{15},\frac{8}{15}$  \\
$21_{\frac{22}{5}}$ & $260.2$ & ${\red {0}},\frac{2}{3},\frac{2}{3},\frac{13}{15},\frac{13}{15},{\red {\frac{1}{5}}},{\red {\frac{13}{15}}},\frac{8}{15},\frac{8}{15},{\red {0}},\frac{2}{3},\frac{2}{3},\frac{4}{15},\frac{4}{15},{\red {\frac{3}{5}}},\frac{1}{3},\frac{1}{3},{\red {\frac{2}{3}}},\frac{13}{15},\frac{13}{15},{\red {\frac{1}{5}}}$  \\
\hline
\end{tabular}
}

~

\centerline{
\def\arraystretch{1.25} \setlength\tabcolsep{3pt}
\begin{tabular}{ |c|c|l|l| }
\hline
$N_{c}$ & $D^2$ & $s_1,s_2,\cdots$ \hfill \blue{$d_1,d_2,\cdots$} \\ 
\hline
$7_{1}$ & $93.25$ & ${\red {0}},{\red {\frac{1}{2}}},{\red {\frac{1}{2}}},{\red {\frac{1}{4}}},{\red {\frac{1}{4}}},{\red {\frac{5}{8}}},{\red {0}}$ \hfill \blue{$1,\zeta_{6}^{2},\zeta_{6}^{2},2+\sqrt{2},2+\sqrt{2},2\zeta_{6}^{2},3+\sqrt{8}$} \\
$14_{0}$ & $186.5$ & ${\red {0}},\frac{3}{4},\frac{1}{4},\frac{1}{4},{\red {\frac{1}{2}}},{\red {\frac{1}{2}}},0,0,{\red {\frac{1}{4}}},{\red {\frac{1}{4}}},\frac{3}{8},{\red {\frac{5}{8}}},{\red {0}},\frac{3}{4}$  \\
$14_{2}$ & $186.5$ & ${\red {0}},\frac{1}{4},\frac{3}{4},\frac{3}{4},{\red {\frac{1}{2}}},{\red {\frac{1}{2}}},{\red {\frac{1}{4}}},{\red {\frac{1}{4}}},\frac{1}{2},\frac{1}{2},\frac{7}{8},{\red {\frac{5}{8}}},{\red {0}},\frac{1}{4}$  \\
$21_{7}$ & $279.7$ & ${\red {0}},\frac{2}{3},\frac{2}{3},\frac{1}{6},\frac{1}{6},\frac{1}{6},\frac{1}{6},{\red {\frac{1}{2}}},{\red {\frac{1}{2}}},\frac{11}{12},\frac{11}{12},\frac{11}{12},\frac{11}{12},{\red {\frac{1}{4}}},{\red {\frac{1}{4}}},\frac{7}{24},\frac{7}{24},{\red {\frac{5}{8}}},{\red {0}},\frac{2}{3},\frac{2}{3}$  \\
$21_{3}$ & $279.7$ & ${\red {0}},\frac{1}{3},\frac{1}{3},\frac{5}{6},\frac{5}{6},\frac{5}{6},\frac{5}{6},{\red {\frac{1}{2}}},{\red {\frac{1}{2}}},{\red {\frac{1}{4}}},{\red {\frac{1}{4}}},\frac{7}{12},\frac{7}{12},\frac{7}{12},\frac{7}{12},\frac{23}{24},\frac{23}{24},{\red {\frac{5}{8}}},{\red {0}},\frac{1}{3},\frac{1}{3}$  \\
\hline
\end{tabular}%
}
\end{widetext}

\vfill

\end{document}